\documentclass[12pt]{article}
\usepackage{epsf,psfig}
\usepackage{amsmath,amssymb}
\usepackage{graphics,graphicx}
\usepackage{supertabular}
\newcommand{\HI}{H\,{\sc i}}

\newcommand{\HeI}{He\,{\sc i}}
\newcommand{\HeII}{He\,{\sc ii}}
\newcommand{\LiI}{Li\,{\sc i}}
\newcommand{\CI}{C\,{\sc i}}
\newcommand{\CII}{C\,{\sc ii}}
\newcommand{\CIII}{C\,{\sc iii}}
\newcommand{\CIV}{C\,{\sc iv}}

\newcommand{\NI}{N\,{\sc i}}
\newcommand{\NII}{N\,{\sc ii}}
\newcommand{\NIII}{N\,{\sc iii}}
\newcommand{\NIV}{N\,{\sc iv}}
\newcommand{\NV}{N\,{\sc v}}
\newcommand{\OI}{O\,{\sc i}}
\newcommand{\OII}{O\,{\sc ii}}
\newcommand{\OIII}{O\,{\sc iii}}
\newcommand{\OIV}{O\,{\sc iv}}

\newcommand{\OVI}{O\,{\sc vi}}
\newcommand{\FII}{F\,{\sc ii}}
\newcommand{\FIV}{F\,{\sc iv}}

\newcommand{\NeII}{Ne\,{\sc ii}}
\newcommand{\NeIII}{Ne\,{\sc iii}}
\newcommand{\NeIV}{Ne\,{\sc iv}}
\newcommand{\NeV}{Ne\,{\sc v}}

\newcommand{\NaIV}{Na\,{\sc iv}}

\newcommand{\MgI}{Mg\,{\sc i}}

\newcommand{\MgVI}{Mg\,{\sc vi}}
\newcommand{\SiI}{Si\,{\sc i}}
\newcommand{\SiII}{Si\,{\sc ii}}
\newcommand{\SiIII}{Si\,{\sc iii}}

\newcommand{\SiVI}{Si\,{\sc vi}}
\newcommand{\SiVII}{Si\,{\sc vii}}
\newcommand{\PII}{P\,{\sc ii}}

\newcommand{\SII}{S\,{\sc ii}}
\newcommand{\SIII}{S\,{\sc iii}}

\newcommand{\ClII}{Cl\,{\sc ii}}
\newcommand{\ClIII}{Cl\,{\sc iii}}
\newcommand{\ClIV}{Cl\,{\sc iv}}

\newcommand{\ArII}{Ar\,{\sc ii}}
\newcommand{\ArIII}{Ar\,{\sc iii}}
\newcommand{\ArIV}{Ar\,{\sc iv}}
\newcommand{\ArV}{Ar\,{\sc v}}

\newcommand{\KIV}{K\,{\sc iv}}
\newcommand{\KV}{K\,{\sc v}}
\newcommand{\KVI}{K\,{\sc vi}}

\newcommand{\CaII}{Ca\,{\sc ii}}

\newcommand{\CaV}{Ca\,{\sc v}}
\newcommand{\CaVI}{Ca\,{\sc vi}}
\newcommand{\VI}{V\,{\sc i}}
\newcommand{\CrII}{Cr\,{\sc ii}}
\newcommand{\CrIII}{Cr\,{\sc iii}}

\newcommand{\MnII}{Mn\,{\sc ii}}
\newcommand{\MnIII}{Mn\,{\sc iii}}
\newcommand{\MnIV}{Mn\,{\sc iv}}
\newcommand{\MnV}{Mn\,{\sc v}}

\newcommand{\FeII}{Fe\,{\sc ii}}
\newcommand{\FeIII}{Fe\,{\sc iii}}
\newcommand{\FeIV}{Fe\,{\sc iv}}
\newcommand{\FeV}{Fe\,{\sc v}}
\newcommand{\FeVI}{Fe\,{\sc vi}}
\newcommand{\FeVII}{Fe\,{\sc vii}}
\newcommand{\CoI}{Co\,{\sc i}}

\newcommand{\NiII}{Ni\,{\sc ii}}
\newcommand{\NiIII}{Ni\,{\sc iii}}

\newcommand{\NiV}{Ni\,{\sc v}}

\newcommand{\BrIII}{Br\,{\sc iii}}
\newcommand{\KrIV}{Kr\,{\sc iv}}
\newcommand{\RbV}{Rb\,{\sc v}}
\newcommand{\BaII}{Ba\,{\sc ii}}
\def\reference{\parskip 0pt\par\noindent\hangindent 0.5 truecm}

\begin{document}
%
%
\title{The Internal Extinction Curve of NGC~6302 and its 
Extraordinary Spectrum}
%


\author{Brent Groves, $^{1}$
   Michael A. Dopita, $^{1}$
   Robert E. Williams, $^{2}$ \and
   Chon-Trung Hua $^{3}$
} 

\date{}
\maketitle

{\center
$^1$ Research School of Astronomy and Astrophysics, Australian
National University, Cotter Rd, Weston, ACT 2611, Australia
\\bgroves@mso.anu.edu.au, Michael.Dopita@anu.edu.au\\[3mm]
$^2$ Space Telescope Science Institute,3700 San Martin Drive, Baltimore, MD,
21218, USA\\wms@stsci.edu\\[3mm]
$^3$ Laboratorie d'Astronomie Spatiale, Marseille, France\\Trung.Hua@astrsp-mrs.fr\\[3mm]
}

%
\begin{abstract}
In this paper we present a new method for obtaining the optical
wavelength-dependent reddening function of planetary nebulae,
using the nebular and stellar continuum. The data used was a spectrum
of NGC~6302 obtained using the Double Beam Spectrograph on the 2.3m
telescope at Siding Springs Observatory over three nights. This resulted
in a spectrum covering a wavelength range $3300-8600$~\AA\, with
a large dynamical range and a mean signal to noise of $>10^2$~\AA$^{-1}$
in the nebular continuum. With such a high S/N the continuum
can be accurately compared with a theoretical model nebular plus
stellar continuum. The nebular electron temperature and density used
in the model are determined using ratios of prominent emission lines.
The reddening function can then be obtained from the ratio of the
theoretical and the observed continuum. In the case of NGC 6302, it is
known that much of the reddening arises from dust within or around the
nebula, so that any differences between the measured reddening law and the
`standard' interstellar reddening law will reflect differences in the
nebular grain size distribution or composition. We find that for NGC 6302,
the visible to IR extinction  law is indistinguishable from `standard'
interstellar reddening, but that the UV extinction curve is much steeper
than normal, suggesting that more small dust grains had been ejected into
the nebula by the PN central star. We have detected the continuum from
the central star and determined its Zanstra Temperature to be of order
150,000K. Finally, using the extinction law that we have determined, we present
a complete de--reddened line list of nearly 600 emission lines, and report
on the detection of the He(2-10) and He(2-8) Raman Features at
$\lambda4331$~\AA\ and $\lambda4852$~\AA, and the detection of
Raman-Scattered \OVI features at 6830 and 7087 \AA. We believe this to be
the first detection of this process in a PN.

\end{abstract}


\bigskip

%
%
\section{Introduction}
Because interstellar dust grains are very small, typically less than a
micron in diameter, their absorption and scattering properties 
are not only composition--dependent but also wavelength--dependent.
Blue and UV light is usually preferentially scattered compared to that of
longer wavelengths. Dust grains can not only absorb and scatter light from
objects, they can also re-emit in the thermal infra-red, polarize light through
grain alignment mechanisms or be accelerated, heated and photoelectrically
charged by the electromagnetic radiation which impinges upon them.

All of these processes are known to occur in the planetary nebula (PN)
environment. In particular, during the Asymptotic Giant Branch (AGB) phase
of evolution, mass--loss releases material into the circum--stellar
environment which has undergone partial nuclear processing in the central
star. Since this environment is fairly cool, dust may be formed by
direct condensation out of the gaseous phase whenever the kinetic
temperature of the gas falls below a critical value which allows solids
to form. In this case, we have a gas which is slowly cooling from higher
temperatures and in which the pressure and supersaturation are high enough
to allow both nucleation and grain growth. However, it is unlikely that there
exists a state of thermodynamic equilibrium in the dust-forming gas, and
shock heating and cooling are often both important. Therefore,
a complex and detailed time-dependent description of the chemical reactions,
usually referred to as a kinetic model, is needed to describe this situation.

Because of the physics of the condensation process, and the interaction
between the grains formed in the flow and the radiation field of the star,
there is a complex relationship between the nature of the grains, their size
distribution and the terminal velocity of the dusty outflow. Kozasa \& Sogawa
(1997) showed that the grain size increases as the mass--loss rate increases,
since the size of the grain produced by condensation depends upon the gas
density in the wind where a strong supersaturation exists in the gaseous
phase and upon the period during which the condensation timescale is much
shorter than the dynamical expansion timescale. On the other hand,
radiation pressure acting upon the grains accelerates the stellar mass-loss
flow (thereby arresting the condensation process). This has been seen
observationally by Loup {\it et al.} (1993) and explained theoretically by
Habing {\it et al.} (1994). The expansion velocities of the carbon rich objects
are larger than those of the oxygen rich AGB stars, and radiation pressure
induced expansion of the atmosphere may limit the size of the typical
carbon-bearing grain to $\sim 50$~\AA, similar to that which is needed to
explain the 2175~\AA\ bump in the interstellar extinction curve. During
the PN phase of evolution, we expect the grain size distribution to be
further modified by radiative destruction processes (photoevaporation and
coulomb destruction by excessive photoelectric charging) and by mechanical
processes (grain coagulation and shattering).

Taking all of these considerations into account, it is clear that we should
expect that the dust formed in the gas ejected during the AGB, and later
observed in PN phase, would be quite unlike like that seen in the
interstellar medium as a whole. It is therefore of great interest to either
observe this dust directly through IR observations, or else through the
extinction produced by it in the optical and UV.

As far as direct observations are concerned, enormous progress has
recently been made using the ISO satellite to obtain spectroscopy of the
far--IR emission features characteristic of different grain materials (Waters
{\it et al.} 1996). The bright southern PN NGC 6302 is an ideal object
for such studies, as it is known to have within it a dense circumstellar
torus containing the bulk of the dust mass (Lester \& Dinerstein, 1984), and
within this, a dense ring of ionised gas, inclined at about 45 degrees to the
plane of the sky (Rodriguez {\it et al.} 1985).
Recently, Kemper {\it et al.} 2002 have reported the detection of features
in the far--IR spectrum of this object which may be ascribed to the silicates
amorphous olivine, forsterite, clino-enstatite, and diopside. In addition
features due to water ice and metallic iron are seen. Remarkably, the
carbonates calcite and dolomite were also detected.

At optical wavelengths, the lack of strong spectral features renders such
exquisite mineralogy impossible. However, because dust grain dimensions
are often comparable to or smaller than the wavelength of light, the dust
extinction curve can in principle be used as a powerful constraint on the grain
size distribution in the nebula.

For PNe we usually characterise the reddening by a single `reddening
constant', $c$, and then assume that the absorption through the optical
wavelength region can be fit by a `standard' Whitford (1958) reddening law.
This curve, f($\lambda$), can then be used to deredden the observed emission
line fluxes. The relationship between the corrected flux, F$_c$, and that
observed, F$_o$ is:

\begin{equation}
\text{F}_c=\text{F}_o\times10^{cf(\lambda)}
\end{equation}

The reddening constant is usually determined from a comparison of the
ratio of the intensities of the Balmer lines, since this `Balmer decrement'
is only slightly dependent upon the temperature and density of the nebula,
and the theoretical values are well--determined. Alternatively, we can
compare the radio continuum flux density and the H$\beta$ flux. The radio
emission is basically free from interstellar reddening and the ratio
between the radio continuum flux and the H$\beta$ flux is determined only
by the electron temperature and the relative helium abundance. A third
technique is to measure the ratio of two emission lines which share a common
upper energy level, such as H$\beta$ and Br$\gamma$ (Ashley, 1990).

All of these methods have their problems. In the first case, the reddening
is determined at only a few discrete wavelengths, and over a restricted
wavelength range. In the other two cases, we may be seeing regions of
ionized gas in the radio or at IR wavelengths which are entirely
dust--obscured in the optical, and therefore we can neither correctly evaluate
the effective total obscuration nor the differential extinction at
different optical wavelengths in the nebular gas.

The motivation behind the work described in this paper, is to obtain an
intrinsic reddening function which does not depend on the Whitford curve,
which is continuous in its wavelength coverage, and which can be used to
place constraints on the grain size distribution in a planetary nebula.

To do this, we have obtained very high signal to noise
observations of NGC 6302 covering the wavelength range $3300--8600$~\AA,
allowing observations of both Paschen and Balmer lines, and of both the
Balmer and the Paschen discontinuities of Hydrogen.  We have then compared
the observed continuum spectral energy distribution to a theoretical
(nebular $+$ stellar) spectral energy template to derive the
reddening function. As far as we are aware, this represents the first
practical application of this novel technique in the literature.

\section{Observations and Reduction}
\label{obs}
NGC 6302 is a very bright, nearby Type I planetary nebula which displays a
bipolar, filamentary structure. Its central star of the nebula is believed
to be very hot, with a temperature possibly as great as 430000 K (Ashley
\& Hyland, 1988). However, the central star has never been identified
either owing to heavy obscuration in the central parts of the nebula, or else
owing to its extreme temperature.

To observe NGC 6302 we used the Double Beam Spectrograph (DBS) (Rodgers, Conroy
\& Bloxham 1988) with its EEV CCD
detectors on the 2.3m telescope Siding Springs Observatory.
A 1200 l/mm grating was used in both the red and the blue arms giving a
wavelength coverage of just over 1000 \AA . We observed NGC 6302 over
three photometric nights (6--8 July 1999) in 6 independent wavelength ranges,
corresponding to three grating settings per arm. These settings, given below,
were chosen to allow a slight overlap between each spectrum, and to avoid
placing strong emission lines in the overlap region:

\begin{tabular}{c c c c}
  1 & 3300-4300 (B) & 5800-6800 (R) & Dichroic \# 1\\
  2 & 4200-5200 (B) & 6700-7700 (R) & Dichroic \# 5\\
  3 & 5100-6100 (B) & 7600-8600 (R) & Dichroic \# 5\\
\end{tabular}

Dichroic filter \#1 is the only one with satisfactory performance in the
UV below the Balmer Discontinuity, but for the second and third grating
setups, Dichroic \#5 was used, since this gives smoother transmission
characteristics in the red.

In order to obtain spectra of great dynamical range, we had to make a series
of exposures of different lengths to ensure that we had good relative
photometry for the bright emission lines such as the [\OIII] $\lambda 5007$
line, which were saturated on the detector in the longer exposures.
Three independent frames were taken for each exposure time to
eliminate cosmic ray events and to reduce the noise in the final spectrum.
The full set of exposures for each pair of grating settings were:
1) 20s, 60s, 200s, 500s, and 1500s; 2) 20s, 60s, 180s, 600s, and
1500s; 3) 500s, and 1500s. Only two exposures were required at the third
setting because of the lack of strong emission lines in these two regions.
Each spectrum was 1850 pixels in length and covered 200 spatial pixels,
each corresponding to 0.91 arc sec. on the sky.

\begin{figure}
{\center
\includegraphics[scale=0.6]{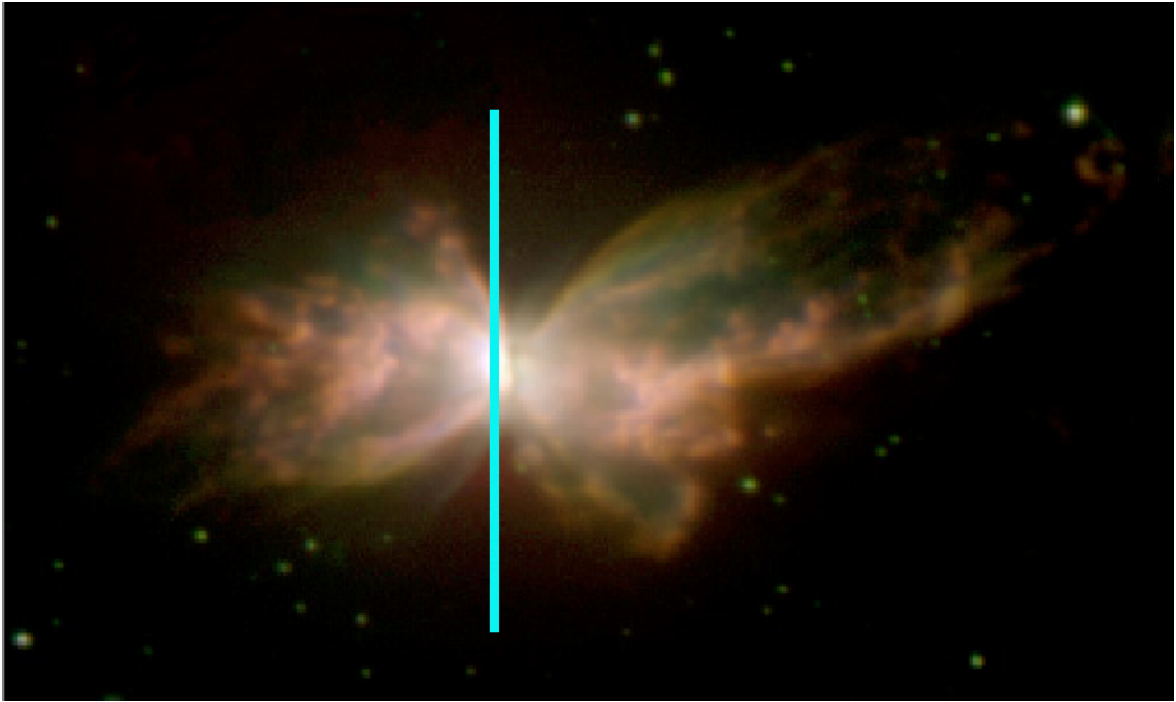}
\caption{The planetary nebula NGC 6302 (Hua, Dopita \& Martinis 1997). The
placement of the 2.3m slit on the nebula is indicated by the blue line. }
\label{NGC6302im}
}
\end{figure}

The slit width was chosen to be 2 arc sec. This optimises the throughput
without appreciable degradation of the spectral resolution. The slit was
placed on the brightest optical region of NGC 6302 as shown in
Fig. \ref{NGC6302im}. This image was obtained using the 2.3m imager, and
is a colour composite of three observations through narrow-band filters
isolating respectively: blue, [\OIII] $\lambda 5007$ ; green, H $\alpha$,
and red, [N~II] $\lambda 6584$.

For wavelength calibration a Neon--Argon arc lamp was used, and for flux
calibration the standard stars EG131 and Feige110 (Bessell, 1999) were
observed. The star EG131 is particularly useful, because it lies not too
far away on the sky from NGC 6302, and can therefore be used as an
atmospheric standard as well. The flat field was generated by observing
through the spectrograph, the diffuse reflection on a white-painted region
of the shutter of the dome of an array of quartz-iodide lamps placed around
the upper secondary support ring structure of the telescope.

The reduction of the data was done using the IRAF package. The reduction
procedure was fairly complex, because of the number of observations, and the
large dynamical range targeted for the final spectrum.

For each frame, the bias observed for that particular night was removed.
However, as the telescope tracks, the spectrograph which is mounted at the
Nasmyth A focus rotates, and consequently there is a temperature shift in
the pre-amplification and CCD control electronics rack which is mounted
on the spectrograph. This results in a systematic bias drift, which has to
be removed in each frame by reference to the bias strip using the tasks
\small{IMSTAT} and \small{IMARITH}. The spectrograph rotation also produces
flexure which results in a small shift of the spectrum. To eliminate this,
each set of three spectra were aligned relative to the observation that
was nearest the arc observation using the IRAF task
\small{IMALIGN) and then combined using the \small{IMCOMBINE} option
with the \small{CCDREJECT} option to remove cosmic ray events.

Flat fields were prepared by dividing each flat field observation by
a low-order spline surface fit to the flat field to remove (to first order)
the gross effects due to the spectral energy distribution of the quartz
iodide lamps. At the UV end of the spectrum, around $3500-3300$\AA, the
accuracy of the flat field is limited by the photon statistics in the
lamp. The flat fielding removes not only the point--like defects due to
dust and blemishes in the CCD, but also the oscillations in the transmission
of the dichroic beam--splitters, which is particularly noticeable in the
red arm, close to the cutoff wavelength.

Following this, one dimensional spectra are extracted from a $\sim 6$
arc sec. length of slit centred on the eastern hot-spot. From this point,
the data reduction follows the standard procedures described in the IRAF
handbooks. The standard stars are also used as atmospheric standards to
correct, as best as possible, for the OH atmospheric molecular band
absorptions in the red.

After reduction, there remain a number of minor but significant problems
in the data. First, the $4200-5200$\AA\ observation suffers from grossly
out of focus ghost images (produced in the camera) of the very bright
[\OIII] $\lambda\lambda 4959,5007$\AA\ lines. These cannot be fully removed
from the data, and corrupt the continuum measurements in the
$\sim 4300-4600$\AA wavelength range. In addition, the [\OIII] lines
themselves were so intense that in the long exposures, not only were the CCD
columns containing these lines completely saturated, but there was also
appreciable bleeding in the line direction as as well, up to the boundary
of the chip near $5300$\AA.  An attempt has been made to correct for this
effect, but the continuum fluxes measured in this spectral range remain
less reliable.

Second, the absolute fluxes measured in the individual spectra
differed from night to night, as judged from the overlap regions.
This is probably mainly due to small errors in the re--positioning of the slit,
despite the fact that the same centering and guide star offset
figures were used for all three nights. Since the first night's observations
cover H$\alpha$ in the red, and H$\delta$ and beyond, down to the Balmer
continuum in the blue, the remaining spectra were normalised by a fixed
multiplicative factor to best remove any discontinuities in the overlap region.

Lastly, spectrograph drifts due to differential flexure problems during the
long exposures provide a larger uncertainty in the absolute
wavelength calibration than is desirable, with errors ranging up to $0.6$\AA.
However, the absolute wavelength scale is generally very accurately determined,
with an error as small as $\sim 0.03$\AA. However, in some cases, the lack
of arc lines in the overlap region sometimes means that the systematic
wavelength error in these regions may increase to up to $0.6$\AA.
Thus, the absolute wavelengths of individual spectral features should only be
measured relative to nearby known hydrogen or helium lines, observed
at the same time as the feature of interest.

In the long exposure (4500s integration time) combined spectrum, several
of the bright lines were saturated on the CCD, sometimes quite grossly.
The regions of saturation were determined, and the flux in these over--exposed
regions was replaced by that measured in a shorter exposure in which
these lines were not saturated. This gives a full spectrum which has a large
wavelength range, high resolution, large S/N ratio and a very large dynamic
range,  shown in Fig. \ref{unredNGC6302}. Typically the mean signal to noise
in the nebular continuum is $>10^2$ \AA$^{-1}$. Hundreds of emission lines
are visible. The identifications ascribed to these, and their relative
intensities are discussed below. Note the prominent Balmer and Paschen
discontinuities in the nebular continuum.

\begin{figure}[!htbp]
{\center
\includegraphics{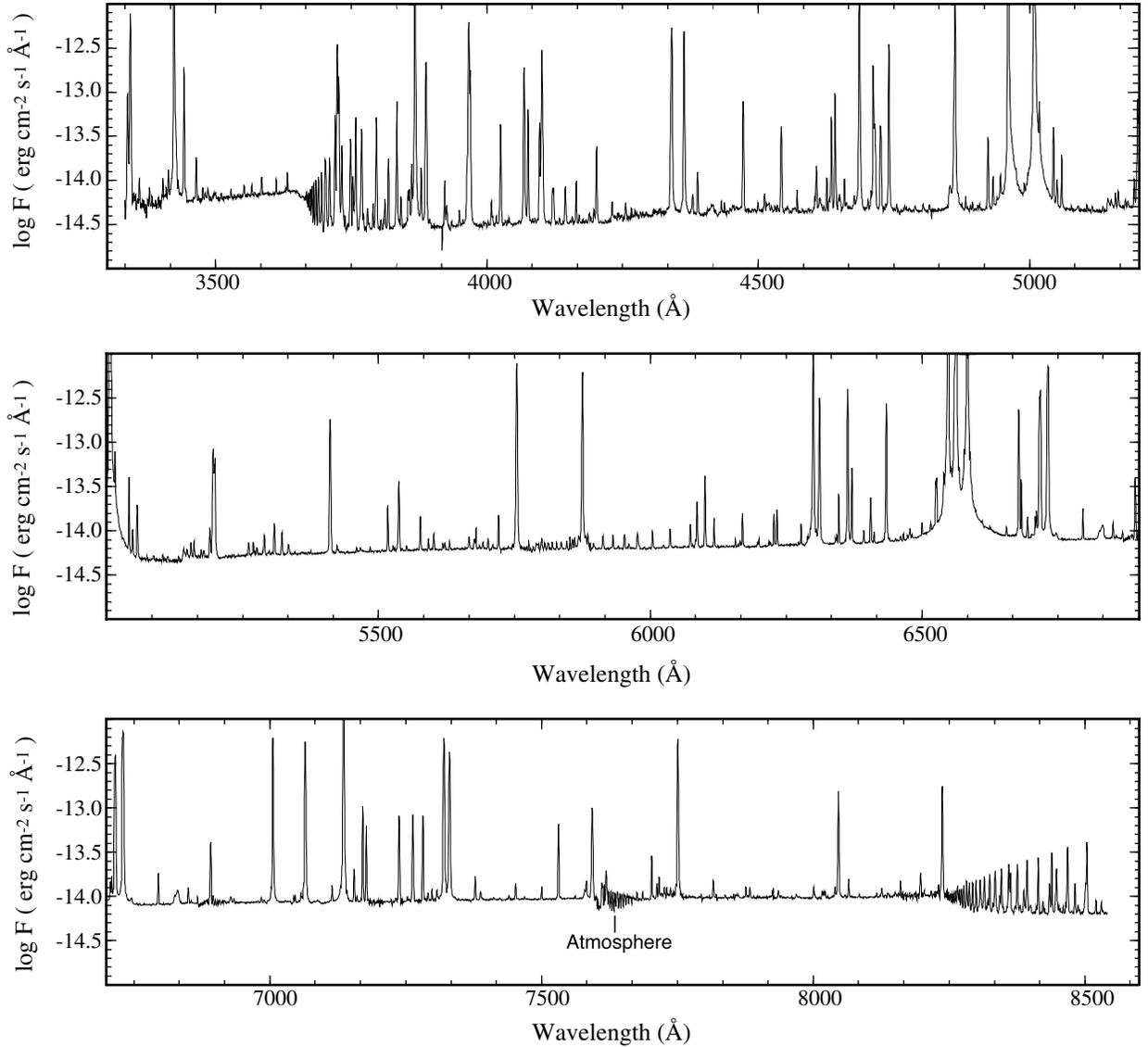}
\caption{ The spectrum of NGC 6302 covering a wavelength range of 3300--8600
\AA\ and a very large dynamic range. Also visible in the spectrum is the
nebular continuum, with a typical S/N greater than$10^{2}$ \AA $^{-1}$. A full
line list from this spectrum is given in in Table \ref{linelist}.}
\label{unredNGC6302}
}
\end{figure}

\section{The Theoretical Nebular Continuum\label{continuum}}
The continuum emission from a planetary nebula comes from several processes.
Given the temperature and density of the nebula and abundance of the emitting
species, the full continuum emission from the nebula can be theoretically
predicted. In the theoretical continuum emission calculated here the three main
nebular emission processes were taken into consideration: free--free emission,
free--bound emission and the two--photon continuum. The two major elements,
Hydrogen and Helium were the only species considered as contributing to these
processes. The theory of these processes, along with useful tables, 
is summarised
by Dopita \& Sutherland (2002).

For the first two processes, a simplified fit to the theoretical continuum
emission due to Hydrogen and the two ionic states of Helium has been
calculated from (Brown \& Mathews 1970), which is applicable over the range
of temperatures liable to be encountered. We fit the peaks of emission
coefficient near prominent discontinuities (such as at the Balmer jump) with
functions of the form:

\begin{equation}\label{gpeak}
\gamma_{\nu}\propto\text{T}^{\alpha}(1+\beta[\text{ln(T})]^2)
\end{equation}

with the constants $\alpha$ and $\beta$ and the constant of proportionality
determined from the data for each peak.

Between discontinuities we fit the gradient of log$(\gamma_{\nu})$ vs. $\nu$
with a power law:

\begin{equation}
\text{gradient}\propto\text{T}^{\alpha}
\end{equation}

with the constants $\alpha$ and the constant of proportionality
determined from the data for each wavelength regions between peaks.

To these we add the theoretical emission from the two-photon process.
The usual assumption, that there is a large optical depth for Ly$\alpha$
was adopted. Finally, the possibility that there is a continuum due
to the central star should also be allowed for in making a fit to the
observed nebular continuum. Here, we simply assume that since the
central star is so hot, the spectrum can be fit by the Rayleigh--Jeans
approximation for a Black Body.

The nebular continuum is normalised to the emissivity of H$\beta$
which can be obtained from Osterbrock (1989) or, equivalently from the
tables in the appendices of Dopita \& Sutherland (2002).

In order to fit this model continuum to the observed data, we need
to observationally determine the parameters of nebular temperature,
nebular density, and abundances by number of the He$^{+}$ and
He$^{++}$ ions relative to the H$^{+}$ ion. These can be obtained
with sufficient accuracy from the normal nebular diagnostics, provided
that an initial estimate of the reddening can be determined, as shown in
the following section.

\section{Determination of Nebular Properties}

There are a number of density sensitive line ratios available in the spectrum.
These include the usual [\OII] $\lambda3726/\lambda3729$ and [\SII]
$\lambda6731/\lambda6717$ line ratios, as well as the [\ArIV]
$\lambda4740/\lambda4711$ and [\ClIII] $\lambda5537/\lambda5517$ line ratios
which are not as frequently used because the lines are fainter, and their use
requires spectra of higher signal to noise. Since all these lines pairs are
close in wavelength, we do not have to worry about reddening corrections.
The densities have been obtained from these ratios using the Australian version
of the MAPPINGS {\sc III} code (Sutherland \& Dopita, 1993). The derived densities are
listed in Table \ref{densities}.

\begin{table}[!htbp]
\centering
\begin{tabular}{|c|c|}
\hline
Line Ratio & Density ($10^4\rm{cm}^{-3}$) \\
\hline
\text{[\OII]}$\lambda3726/\lambda3729$ & 0.5 \\
\text{[\SII]}$\lambda6731/\lambda6717$ & 0.8 \\
\text{[\ArIV]}$\lambda4740/\lambda4711$ & 1.3 \\
\text{[\ClIII]}$\lambda5537/\lambda5517$ & 2.0 \\
\hline
\end{tabular}
\caption{\label{densities} The measured electron density for
NGC 6302 from several line ratios}
\end{table}

In order to estimate the nebular temperature from line ratios, we first need
to adopt an estimate of reddening. This was done as a first approximation by
measuring the Balmer Decrement and comparing with the theoretical decrement
for an assumed electron temperature of 15,000K (the choice of the electron
temperature is not critical, since the Balmer decrement is little dependent on
the temperature). Individual line ratios can then be dereddened using the
Whitford reddening curve (as tabulated by Kaler, 1976). We find a reddening
constant of $c=1.2$, in excellent agreement with that determined in the same
way by Aller {\it et al.} 1981. They found $c=1.22$. It is interesting to note
that these reddening values are lower than that obtained either from the
H$\beta$ to Br$\gamma$ ratio ($c=2.44$, Ashley, 1990), or the H$\beta$ to
radio continuum ratio ($c=2.1$, Milne \& Aller, 1985; $c=2.1$, Ashley
\& Hyland, 1988). This is clear evidence that there exists a highly--obscured
inner region in the nebula, visible only in the IR or at radio wavelengths.

The temperature sensitive lines used were [\OIII] $\lambda4363/\lambda5007$,
[\NII] $\lambda5754/\lambda6583$, [\SII] $\lambda4076/\lambda6731$ and
[\OI] $\lambda5577/\lambda6300$. The temperatures obtained from these ratios
are listed in Table \ref{temps}. Since the emission is heavily weighted
towards the high--excitation regions in our spectrum, we adopt a
temperature of $\text{T}_e\sim1.5\times10^4 K$ for the continuum model
fitting described below.

\begin{table}[!htbp]
\centering
\begin{tabular}{|c|c|}
\hline
Line Ratio & Temperature ($10^4 \rm{K}$)\\
\hline
\text{[\OIII]}$\lambda4363/\lambda5007$ & 1.5 \\
\text{[\NII]}$\lambda5754/\lambda6583$ & 1.4 \\
\text{[\SII]}$\lambda4076/\lambda6731$ & 1.2 \\
\text{[\OI]}$\lambda5577/\lambda6300$ & 1.0 \\
\hline
\end{tabular}
\caption{\label{temps} The measured electron temperature for NGC 6302 from
several line ratios}
\end{table}

To determine the helium ionic abundance, ratios between the Pickering (n-4) and
(n-3) lines and the Balmer lines for \HeII\ and singlet lines such as
$\lambda6678$ and the Balmer lines for \HeI\ were used. We use the singlets for
this purpose because, unlike the triplets, they are unaffected by optical
depth and line transfer problems. The flux ratios are then be used with
the data from Osterbrock (1989) and from Dopita \& Sutherland (2002) to
obtain the abundance of the He$^{+}$ and He$^{++}$ions relative to H$^{+}$.

The final parameters used in the calculation of the continuum are, temperature,
$\text{T}_e\sim1.5\times10^4$ K, electron density,
$\text{n}_e\sim1.0 \times 10^4$ cm$^{-3}$,  He$^{+}$ to H$^{+}$ abundance ratio
$\sim0.11$ and He$^{++}$ to H$^{+}$ abundance ratio $\sim0.07$.

\section{The Reddening Curve for NGC 6302 \label{secred}}

With the electron density, temperature and helium ionic abundances estimated
above, we first built a theoretical continuum of NGC 6302, as
shown in Fig. \ref{N6302cont}. This theoretical emission was then
divided by the
spectrum from  NGC 6302 to provide an initial estimate of the
reddening function.
The result can be seen in Fig. \ref{stellar}. The reddening function should be
a smooth curve which is defined by the highest points in this function. The
detailed structure is due to the individual emission lines. However, even
ignoring these emission line features, large steps are evident at both the
Balmer and Paschen jumps. These steps would not be removed even were we to
assume a much larger electron temperature, and in any case the residual
Balmer and the Paschen jumps cannot both be simultaneously removed for any
assumed value of the electron temperature.

We are forced to conclude that these discrepancies are the result of the
presence of a reflected and/or direct continuum from the hot central star as
discussed in \S\ref{continuum}, producing a Rayleigh--Jeans tail of a blackbody
spectrum spectrum in the visible ($F_{\lambda}\propto \lambda^{-4}$).

\begin{figure}[!p]
{\center
\includegraphics[scale=0.65]{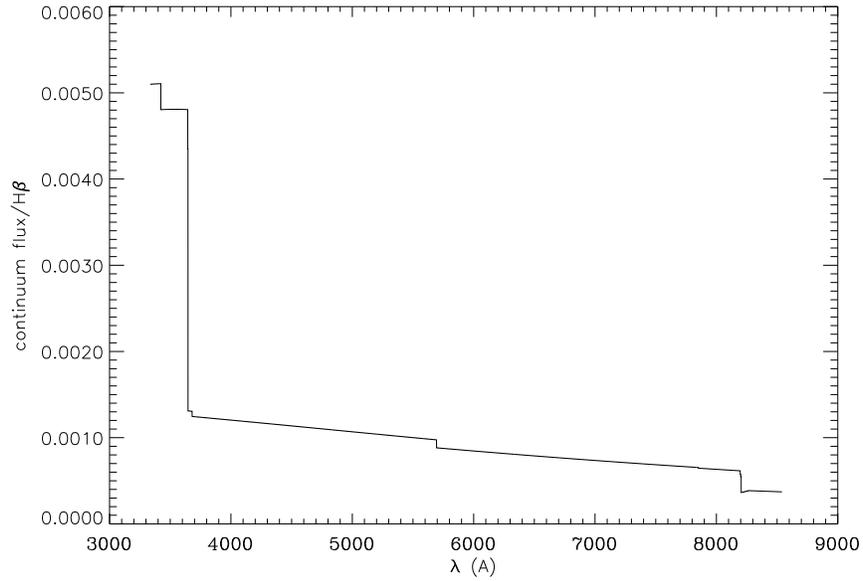}
\caption{The theoretical continuum of NGC~6302 relative to
H$\beta$}
\label{N6302cont}
}
\end{figure}

\begin{figure}[!hp]
{\center
\includegraphics[scale=0.65]{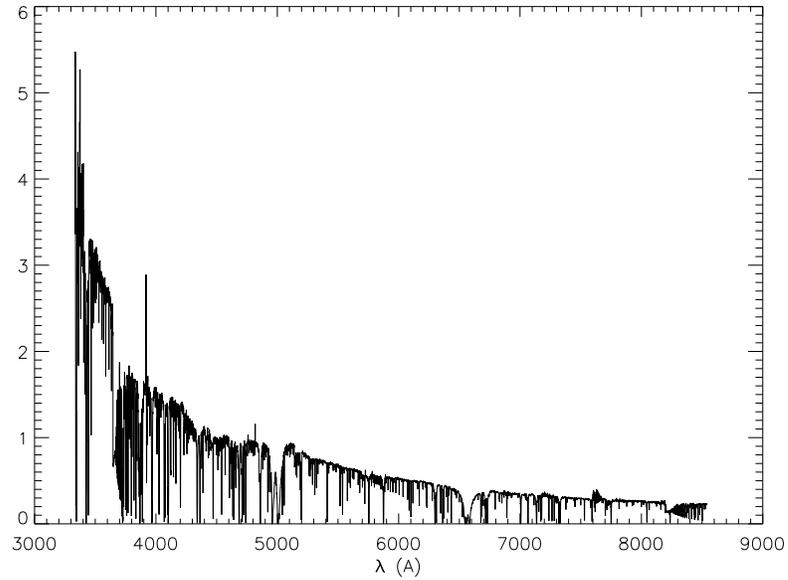}
\caption{The theoretical continuum divided by the spectrum of NGC~6302 showing
the need for addition of a stellar continuum. The emission lines in 
the spectrum
are responsible for the low values in this curve. The apparent features in the
continuum between 4300 and 5200 \AA are artifacts of the saturation effects and
ghost images of the [\OIII] lines described in section \ref{obs} and should be
ignored.}
\label{stellar}
}
\end{figure}

The amount of stellar continuum which we need to add to match the
observed Balmer jump determines the correct scaling factor of this Black--Body
component to add to the continuum template. The correctness of this
scaling factor is evidenced by the fact that, when this is done, the
other jumps such as the Paschen and HeII jumps also match the observations.
The resultant continuum model including the stellar contribution is shown
in Fig. \ref{contstar} and the result of division of this by the observational
data is shown in Fig. \ref{nostellar}.

This represents the first direct detection of the central star of
NGC 6302 by any technique. However, it is more likely that this stellar flux
represents scattered light rather than direct light, since the slit was
displaced from the physical centre of the nebula by more than a slit width,
the nebula is known to be extremely optically thick at its centre
(Ashley, 1990) and direct imaging searches for the central star have so
far failed (Ashley, 1988).

With this determination of the amount of stellar continuum in the spectrum
we can now measure the Zanstra temperature of the central star, assuming that
the stellar continuum is seen directly, rather than being scattered into the
line of sight by the dusty torus, and that the central star lies fully within
the slit.

At $\lambda 4681$, we find that the stellar continuum is $1.79\times 10^{-4}$
that of H$\beta$. This gives  a flux for the central star at $\lambda 4681$ of
F$_* = 6.83 \times 10^{-16}$ erg~cm$^{-2}$ (mW~m$^{-2}$). The global H$\beta$
flux for NGC 6302 is log F$_{\mathrm{H}\beta} = -10.55$ (Perek, 1971), giving
a H$\beta$ to stellar flux ratio,
log $\mathrm{F}_{\mathrm{H}\beta}/\mathrm{F}_{*} =4.61$.  Using the figures
provided in Pottasch (1984) this leads to an estimate for the Zanstra 
temperature
of T$_{z} \sim 1.5 \times 10^5$ K.

This temperature is well below the 430,000 K determined by Ashley \& Hyland
(1988) using high excitation silicon lines. If we are not seeing the central
star directly, but through scattered light, this discrepancy will only
increase. In general, the Zanstra method is known to systematically
underestimate the stellar temperature, as this method assumes a 
blackbody stellar
continuum which usually does not apply to such hot, high gravity 
stars. However,
given that we are almost certainly observing the central star in 
scattered light,
it is quite likely that this star may be a binary with a fairly hot companion.

\begin{figure}[!p]
{\center
\includegraphics[scale=0.65]{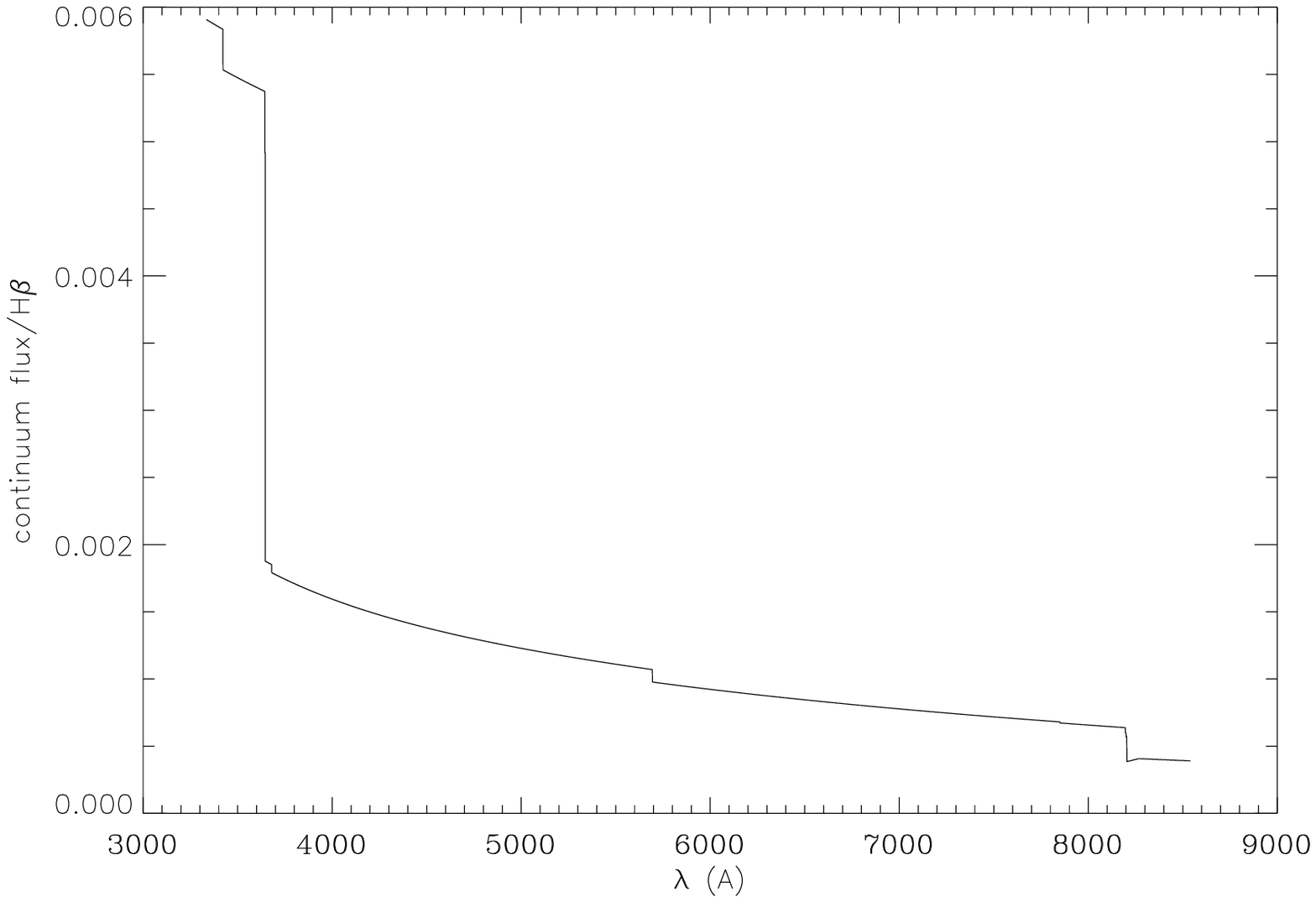}
\caption{The theoretical continuum of NGC6302, including the
direct and reflected stellar continuum, relative to H$\beta$}
\label{contstar}
}
\end{figure}

\begin{figure}[!hp]
{\center
\includegraphics[scale=0.65]{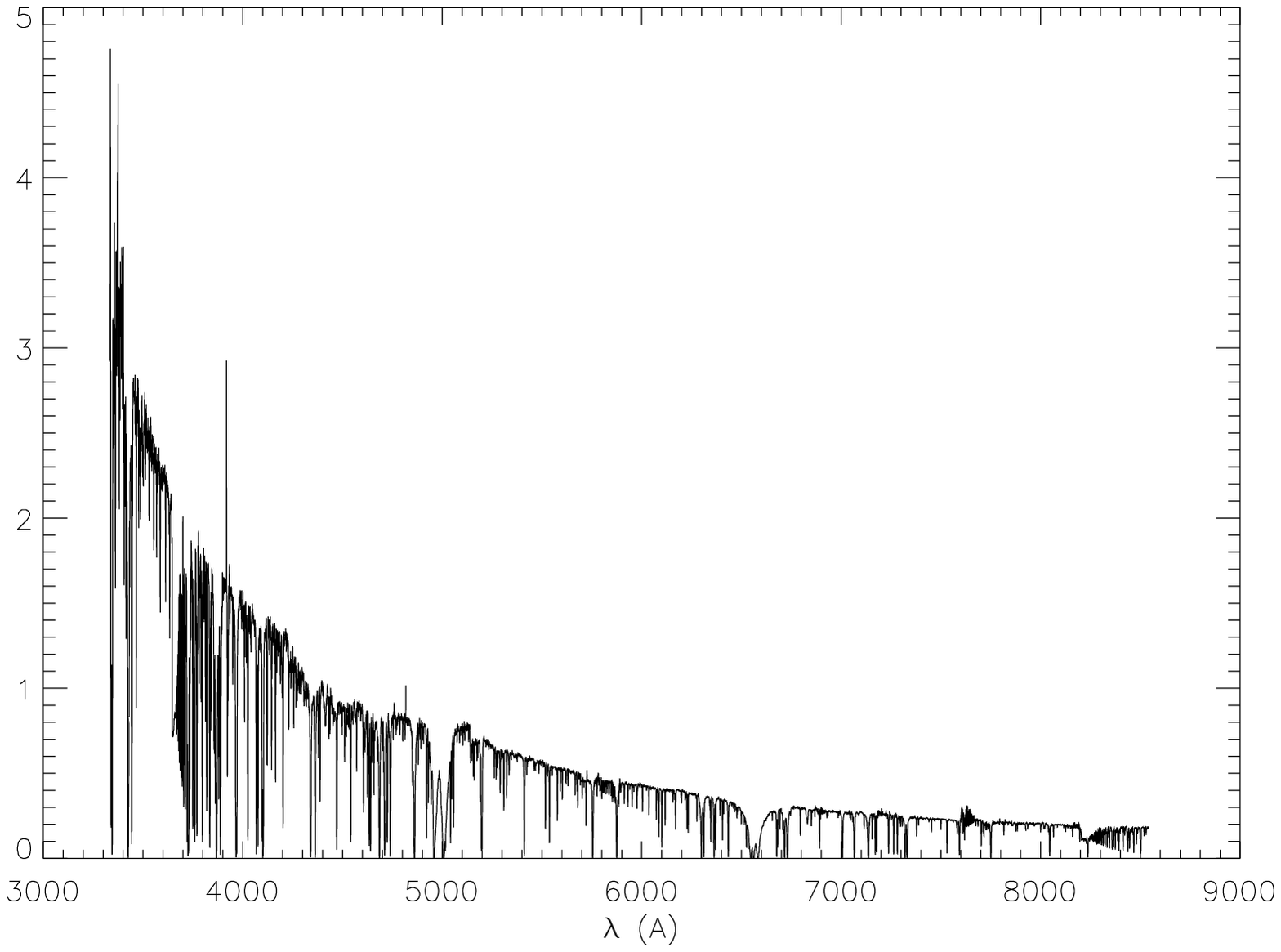}
\caption{The full theoretical continuum, including stellar
contribution, divided by the spectrum of NGC6302, showing the reddening of the
nebula}
\label{nostellar}
}
\end{figure}

With the continuum model fit described above, including the stellar
continuum, the reddening curve was determined from the ratio of the model
continuum to the observed continuum. The reddening curve was fitted 
in IRAF as a
6th. order cubic spline, which osculated the upper envelope of curve of
Fig. \ref{nostellar}. As a comparison, the logarithm of this curve is plotted
against the Whitford reddening curve, with the constant of the Whitford curve
taken to be $c=1.2$. This value provides the best fit to the reddening 
curve, and
also agrees with that previously determined from the Balmer decrement. The
goodness of fit indicates that this method is another way in
which the reddening constant can be calculated.  As can be seen in Fig.
\ref{whitcomp} the curves are remarkably similar, showing that the use of the
Whitford curve for optical measurements of planetary nebulae proves a
remarkably good approximation. However the two curves are systematically
different at shorter wavelengths; the curve for NGC6302 is much
steeper in this region. We can take this as an indication there are
many more small grains along our line of sight to NGC6302 than would be the
case in a typical sightline through the interstellar medium.
These small grains are undoubtedly intrinsic to the nebula, having been
earlier ejected by the central star, and, possibly shattered in their
passage through the nebula by grain--grain collisions.

\begin{figure}[!htbp]
{\center
\includegraphics[scale=0.8]{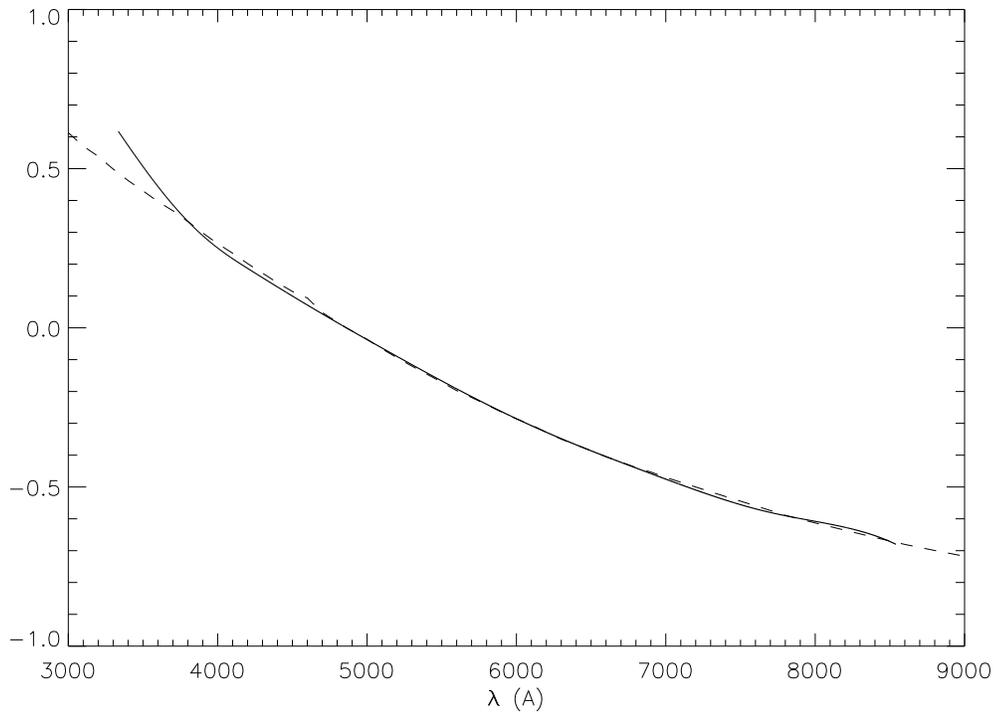}
\caption{ A comparison of the reddening curve for NGC6302 (solid curve) and
the Whitford reddening curve (dashed line) with a logarithmic reddening
constant $c=1.2$}
\label{whitcomp}
}
\end{figure}

\section{The Line Spectrum of NGC6302}
With the reddening curve derived above, the full spectrum of
NGC6302 was dereddened, and the model continuum removed to leave
us with a spectrum containing only the de--reddened emission lines.

The measured wavelengths were then shifted to zero velocity by
using a local fit to the known wavelengths of the hydrogen and helium
recombination lines, or, in the scarcity or absence of these, to
forbidden lines for which very accurate wavelengths are known (\emph{e.g.}
Dopita \& Hua, 1997). The flux and central wavelengths of each emission
feature was then measured using the Gaussian fitting procedure in the IRAF
task \small{SPLOT}, and a line identification was attempted. For this
purpose the earlier spectrum of Aller {\it et. al} (1981), and the very
nice work by Liu {\it et al.} (2000) was very helpful. Extensive use was also
made of the web--based \emph{Atomic line list} v2.04, to be found at
http://www.pa.uky.edu/$\sim$peter/atomic/.

The complete list of identifications, wavelengths, de--reddened line fluxes,
and estimated errors are given for the nearly 600 emission lines detected
in Table \ref{linelist}, below. As mention in \S 2 several of the bright
lines, including [\OIII], H$\alpha$ and [\NII] were grossly oversaturated in
the long exposures. This left only the shortest exposures to measure the flux
in these lines and the percentage errors in the measured fluxes are
consequently larger than the measurement errors of many of the weaker lines.

The spectrum is incredibly rich, and would reward a detailed analysis,
which is not within the scope of this paper. However, a number of
interesting points are worth remarking on here.

First, the He(2-10) and He(2-8) Raman Features are clearly visible at
$\lambda4331$\AA\ and $\lambda4852$\AA. These had only previously
been reported in the PN NGC 7027 P\'equignot {\it et al.} (1997).

In addition, the Raman-Scattered \OVI\ doublet by the enhanced
hydrogenic cross--section near the $3^2$P level gives rise to
velocity--broadened lines at 6830 and 7087 \AA. The theory of this
process was first described by Schmid (1989). The apparent line widths of
8.3 and 9 \AA~FWHM respectively for these lines is an amplification of the
Doppler line width of Ly$\beta$ by a factor of about 6.7. This amplification
is due to the difference in energy between the
incident Ly$\beta$ photon and the outgoing, scattered photons.
For this process to
work we require a very high flux in the \OVI\ doublet at 1032 and 1038 \AA,
in the same region of space where there is a high column density of
neutral hydrogen. Normally these conditions are only encountered
in symbiotic stars, and it is believed that this is the first time these
lines have been detected in a planetary nebula.

Third, the recombination lines of Si (\emph{e.g.} \SiII\ $\lambda3862.60$\AA,
$\lambda5041.0$\AA\ and \SiIII\ $\lambda3956.64$\AA\ are unusually strong.
Taken along with the extraordinary
strength of the $1.96 \mu$m [\SiVI] and $2.486 \mu$m [\SiVII] lines in the
infrared (Ashley \& Hyland, 1988), this is direct evidence that silicaceous
dust is being destroyed in the inner nebula. The smaller grains that
we see here then may have been produced by grain--grain collisions which
have led to grain shattering.

Fourth, in addition to these recombination lines, there is a rich
set of recombination lines of more abundant elements. A crude
analysis of these to estimate the abundances of several ions is shown in Table
\ref{abund}. Though the estimates vary by a factor of two in some ions, they
do not lead one to believe that the abundances derived
from recombination lines are unusually high, unlike the case reported
by Liu {\it et al.} (2000). Note that this observation also supports our assumption for a
high temperature when calculating the continuum emission. The difference in
Balmer jump and [\OIII] temperatures correlates with the difference between the
forbidden line and recombination line abundances (Liu {\it et al.} (2001)), and with the
measured abundances approximately the same, the temperatures should also be similar.

\section{Discussion \& Conclusion}
A high signal-to-noise ratio, high resolution spectrum of the bright
planetary nebula NGC 6302 was obtained with a wavelength range covering the
visible spectrum and its continuum has been used to provide the first detection
of the central star of NGC 6302, and to determine the reddening function
of the dust in the nebula.

As far as the authors know this is the first time the continuum of a planetary
nebula has been measured to such accuracy over such a wide range, and the
first time the intrinsic reddening curve of a nebula been determined from
the form of the nebular continuum. Certainly, the continuum distribution
of planetary nebulae have been used before, but mainly to measure the electron
temperature of the nebulae (Liu \& Danziger 1993).

The UV steepening of the reddening curve of NGC 6302 is taken to mean that
there is a higher abundance of small dust grains in the nebula than is found
in the interstellar medium. However, with only one example, it is not known
whether this property is common to all planetary nebulae or just to those of
Type I composition.

\section*{Acknowledgments}
We wish to acknowledge the use of The Atomic Line List
(http://www.pa.uky.edu/$\sim$peter/atomic/) in the identification of the emission
lines made here.

M. Dopita wishes to thank the Visitor Program of the Space Telescope
Science Institute, and of the Universit{'}e d'Aix--Marseille during
his visit to LAS, during which the spectroscopic analysis described here
was carried through. He would also like to acknowledge the support of the
Australian National University and the Australian Research Council (ARC)
for his ARC Australian Federation Fellowship, and also under ARC Discovery
project DP0208445.

\section*{References}

\reference Aller, L.H., Ross, J.E., O'Mara, B.J. \& Keyes, C.D.
1981, MNRAS, 197, 95.
\reference Ashley, M.C.B. 1988, PhD Thesis, The Australian National University.
\reference Ashley, M.C.B. 1990, PASA, 8, 360.
\reference Ashley, M.C.B \& Hyland, A.R. 1988, ApJ, 331, 532.
\reference Bessell, M.S. 1999, PASP, 111, 1426.
\reference Brown, R. L., \& Mathews, W. G. 1970, ApJ, 160, 939.
\reference Dopita M. A. \& Sutherland, R. S. 2002, \emph{Astrophysics
of the Diffuse Universe}, Springer-Verlag:Berlin, in press.
\reference Fitzpatrick, E. L. 1999, PASP, 111, 63.
\reference Habing, H.J., Tignon, J. \& Tielens, A.G.G.M. 1994,
A\&A, 286, 523.
\reference Hua, C. T., Dopita, M. A. \& Martinis, J. 1997, A\&AS, 133, 361
\reference Kaler, J. B. 1976, ApJS, 31, 517.
\reference Kemper, F., J\"ager, C., Waters, L.B.F.M, Henning, Th.,
Molster, F.J., Barlow, M.J., Lim, T. \& de Koter, A. 2001, Nature,
415, 295.
\reference Kozasa, T. \& Sogawa, H. 1997, Ast. \& Space Sci., 251, 165.
\reference Liu, X. \& Danziger, J. 1993, MNRAS, 263, 256.
\reference Liu, X.-W., Luo, S.-G., Barlow, M.~J., Danziger, I.~J., \& Storey,
P.~J. 2001, MNRAS, 327, 141
\reference Liu, X.-W., Storey, P. J., Barlow, M. J., Danziger, I. J.,  Cohen,
M., \& Bryce, M. 2000, MNRAS, 312, 583.
\reference Loup, C., Foreville, T., Omont, A. \& Paul, J.F. 1993,
A\&A Suppl., 99, 291.
\reference Osterbrock, D. E. 1989, Astrophysics of Gaseous Nebulae
and Active Galactic Nuclei (Mill Valley, CA: University Science Books)
\reference P\'equignot, D., Baluteau, J.-P., Morisset, C. \& Boisson,
C. 1997, A\&A, 323, 217.
\reference Roche, P. F. 1989, in IAU symp. 131, Planetary Nebulae,
ed. S. Torres-Piembert (Dordecht: Reidal), 117
\reference Rodgers, A., Conroy, P. \& Bloxham, G. 1988, PASP, 100, 626
\reference Rodriguez, L.F. {\it et al.} 1985, MNRAS, 215, 353.
\reference Schmid, H.M. 1989, A\&A, 211, 31.
\reference Spitzer, L. \& Greenstein, J. L. 1951, ApJ, 114, 407.
\reference Sutherland, R.S. \& Dopita, M.A. 1993, ApJS, 88, 253.
\reference Waters, L.B.F.M. {\it et al.} 1996, ApJ, 315, L361.
\reference Whitford, A. E. 1958, AJ, 63, 201.

\newpage
\topcaption{Dereddened Line fluxes in the core of NGC6302 relative
to (H$\beta$ = 100.0).
\label{linelist}}
\small
\tablefirsthead{\\ \hline\hline\\
{$\lambda _{\mathrm{0}}$}
&{Ion}
& {$\lambda _{\mathrm{obs}}$}
& {$I(\lambda )$}
& {Error}
& {Comment~~~~~~~~~~~~~~~~~~~~~~~~~~~~~~}\\~\\
\hline \\
}
\tablehead{&\multicolumn{1}{r}{Table 3. -}&{\it Continued.}&\\ \\ \hline\hline\\
{$\lambda _{\mathrm{0}}$}
&{Ion}
& {$\lambda _{\mathrm{obs}}$}
& {$I(\lambda )$}
& {Error}
& {Comment~~~~~~~~~~~~~~~~~~~~~~~~~~~~~~ }\\~\\
\hline \\
}

\begin{supertabular}{llllll}
3340.74 & \OIII & 3340.41 & 19.20 & $\pm 3.0$ & Bowen fluorescent line\\
3345.50 & [\NeV] & 3345.47 & 147.5 & $\pm 15$ & \\
3362.20 & [\NaIV] & 3362.07 & 1.196 & $\pm 0.4$ & \\
& ? & 3381.07 & 0.498 &  & \\
& ? & 3385.23 & 0.219 &  & \\
& ? & 3392.17 & 0.147 &  & \\
3405.74 & \OIII & 3405.55 & 1.125 & $\pm 0.3$ & Bowen fluorescent line \\
& ? & 3411.42 & 0.342 &  & \\
3415.26 & \OIII & 3415.31 & 1.154 &  & Bowen line \\
3425.5 & [\NeV] & 3425.87 & 534.1 & $\pm 20$ & \\
& ? & 3434.02 & 0.380 &  & \\
3444.05 & \OIII& 3444.09 & 34.22 & $\pm 4$ & Bowen fluorescent line \\
3466.5 & [\NI] & 3466.61 & 2.118 &  & \\
3467.54 & \HeI &  &  &  & \\
3478.97 & \HeI & 3478.75 & 0.405 & $\pm 0.04$ & \\
3483.38 & \NII & 3483.19 & 0.380 & $\pm 0.06$ & Blend 3.7\AA\ FWHM \\
3487.73	& {\HeI} (42) + & 3488.18 & 0.605 & $\pm 0.08$ & Blend 2.9\AA\ FWHM \\
3488.7	& [\MgVI] &  &  &  & \\
3498.64 & {\HeI} (40) & 3498.56 & 0.299 & $\pm 0.016$ & Blend  \\
3502.36 & {\HeI} + & 3502.01 & 0.305 &  &  \\
3502.0 & [\MgVI] &  &  &  & \\
3512.51 & {\HeI} (38) & 3512.48 & 0.183 &  &  \\
3530.49	& {\HeI} (36) & 3530.55 & 0.222 &  &  \\
3554.40 & {\HeI} (34) & 3554.46 & 0.403 & $\pm 0.025$ &  \\
3568.5 & \NeII & 3568.55 & 0.329 & $\pm 0.02$ &  \\
3574.6 & \NeII & 3574.49 & 0.083 & $\pm 0.006$ &  \\
3587.26 & {\HeI} (31) & 3587.04 & 0.640 & $\pm 0.025$ &  \\
3613.64 & {\HeI} (6) & 3613.70 & 0.528 &  &  \\
3631.3 & \SiIII & 3631.28 & 0.151 &  &  \\
3634.24 & {\HeI} (28) & 3634.36 & 0.718 & $\pm 0.012$ &  \\
3671.47 & {\HI} H24 & 3671.52 &  &  &  \\
3673.76 & {\HI} H23 & 3673.82 &  &  &  \\
3676.36 & {\HI} H22 & 3676.41 &  &  &  \\
3679.35 & {\HI} H21 & 3679.40 & 0.787 & $\pm 0.05$ &  \\
3682.81 & {\HI} H20 & 3682.85 & 0.931 & $\pm 0.05$ &  \\
3686.83 & {\HI} H19 & 3686.83 & 1.043  & $\pm 0.05$ &  \\
3691.55 & {\HI} H18 & 3691.59 & 1.173 & $\pm 0.05$ &  \\
3694.21 & \NeII & 3694.98 & 0.123 &  &  \\
3697.15 & {\HI} H17 & 3697.14 & 1.397 & $\pm 0.05$ &  \\
3703.85 & {\HI} H16 + & 3704.26 & 2.845 & $\pm 0.05$ & Blend 2.4\AA\ FWHM \\
3705.00	& {\HeI} & 3704.26 &  &  &  \\
3711.97 & {\HI} H15 & 3711.98 & 2.061 & $\pm 0.05$ &  \\
3715.16 & {\HeII} (4-29) & 3715.25 & 0.161 &  &  \\
3717.2 & \SiII & 3717.75 & 0.096 &  &  \\
3721.63 & [\SIII] + & 3721.76 & 7.271 & $\pm 0.17$ &  \\
3721.94 & {\HI} H14 + &  &  &  & \\
3720.40 & {\HeII} (4-28) & 3721.78 & 7.319 &  & \\
3726.03 & [\OII] + & 3726.04 & 43.65 & $\pm 3.0$ &  \\
3726.26 & {\HeII} (4-27) &  &  &  & \\
3728.81 & [\OII] & 3728.71 & 19.58 & $\pm 1.7$ &  \\
3734.37 & {\HI} H13 + & 3734.32 & 3.136 & $\pm 0.11$ &  \\
3732.83 & {\HeII} (4-26) &  &  &  & \\
3736.81 & \OIV & 3737.20 & 0.160 &  &  \\
3740.22 & {\HeII} (4-25) & 3740.17 & 0.187 & $\pm 0.01$ &  \\
& ? & 3743.31 & 0.037 &  &  \\
3748.60 & {\HeII} + (4-24) &  &  &  & \\
3750.15 & {\HI} H12 + & 3750.13 & 3.494 & $\pm 0.12$ &  \\
lamda & {\HeI} (24) &  &  &  & \\
3754.69 & {\NIII} + & 3754.83 & 1.196 &  $\pm 0.09$ &  \\
3754.70 & \OIII &  &  &  & \\
3759.88 & {\OIII} + & 3759.83 & 6.255 &  $\pm 0.22$ &  \\
3758.14 & {\HeII} (4-23) &  &  &  & \\
3770.63 & {\HI} H11 + & 3770.62 & 4.424 &  $\pm 0.10$ &  \\
3770.73 & {\HeI} + &  &  &  & \\
3769.07	& {\HeII} (4-22) &  &  &  & \\
3774.02 & \OIII & 3774.12 & 0.155 &  & \\
3777.42 & \OII & 3777.32 & 0.033 &  & \\
3781.68 & {\HeII} (4-21) & 3781.76 & 0.274 & $\pm 0.02$ & \\
3784.89 & \HeI & 3784.63 & 0.070 & $\pm 0.005$ & \\
3791.28 & \OIII & 3791.33 & 0.316 & $\pm 0.01$ & \\
3796.33	& {\HeII} (4-20) + & 3797.88 & 5.872 & $\pm 0.20$ & \\
3797.90 & {\HI} H10  &  &  &  & \\
3805.78	& \HeI & 3806.04 & 0.110 & $\pm 0.01$ & \\
3811 & O VI? & 3811.29 & 0.044 &  &\\
3813.49 & {\HeII} (4-19) + & 3813.50 & 0.399 & $\pm 0.01$ & \\
3813.54 & [\FeVI]  &  &  &  & \\
3819.61 & {\HeI} & 3819.93 & 1.671 &  & \\
3833.80 & {\HeII} (4-18) + & 3835.37 & 8.774 &  & \\
3835.38 & {\HI} H9 &  &  &  & \\
3839.79 & [\NiV] & 3839.96 & 0.108 &  & \\
3842.81 & \OII & 3843.07 & 0.427 &  & \\
3853.7 & \SiII & 3853.89 & 0.079 & $\pm 0.01$ & \\
3856.02 & {\SiII} + & 3856.05 & 0.506 &  & \\
3856.59 & {\SiII} + &  &  &  & \\
3856.13 & \OII &  &  &  & \\
3858.07 & {\HeII} (4-17) & 3858.09 & 0.559 &  & \\
3862.60 & \SiII & 3862.77 & 1.667 & $\pm 0.13$ & \\
3869.06 & [\NeIII] + & 3868.76 & 210.8 & $\pm 6.0$ & \\
3867.48 & {\HeI} &  &  &  & \\
& ? & 3880.20 & 1.412 &  & \\
3887.44 & {\HeII} (4-16) + & 3888.82 & 23.89 & $\pm 0.35$ & \\
3888.64	& {\HeI} + &  &  &  & \\
3889.05 & {\HI} H8 &  &  &  & \\
& ? & 3895.3 & 0.144 & $\pm 0.05$ & Blend 4.6\AA\ FWHM \\
3923.48 & {\HeII} (4-15) & 3923.44 & 0.698 & $\pm 0.03$ & \\
3926.55 & \HeI & 3926.85 & 0.298 & $\pm 0.03$ & \\
3933.66 & \CaII & 3933.52 & & & seen in Absorption \\
3935.95 & \HeI & 3936.07 & 0.047 &  & \\
3950.31 & [\NiIII] & 3950.42 & 0.174 & $\pm 0.015$ & \\
3956.64 & \SiIII & 3956.62 & 0.064 &  & \\
3964.73 & \HeI & 3964.80 & 1.300 & & Uncertain: on wing of [\NeIII] line \\
3967.79 & [\NeIII] + & 3967.44 & 59.50 & $\pm 1.70$ & \\
3968.43 & {\HeII} (4-14) &  &  &  & \\
3970.07 & {\HI} H7 & 3970.12 & 16.00 &  & \\
3994.62 & [\FeVI] + & 3994.80 & 0.056 & $\pm 0.008$ & Blend 2.2\AA\ FWHM \\
3994.99	& \NII &  &  &  & \\
3997.88 & [\CaV] + & 3997.97 & 0.055 & $\pm 0.008$ & Blend 3.1\AA\ FWHM \\
3998.63 & \NIII &  &  &  & \\
4003.58 & \NIII & 4003.33 & 0.038 &  & \\
4009.25 & \HeI & 4009.25 & 0.292 &  & \\
4011.1 & \NI + & 4011.27 & 0.070 &  & \\
4011.6 & \CIII &  &  &  & \\
4018.1 & \NII & 4018.17 & 0.086 & $\pm 0.007$  & \\
4025.6 & {\HeII} (4-13) + & 4025.98 & 4.102 & $\pm 0.11$ & \\
4026.19 & \HeI &  &  &  & \\
4035.08 & \NI & 4034.72 & 0.052 & $\pm 0.005$  & Blend 2.8\AA\ FWHM \\
4041.31 & \NII & 4041.31 & 0.070 & $\pm 0.013$  & \\
4043.53 & \NII & 4043.56 & 0.028 &  & \\
4060.2 & [\FIV] +  & 4060.22 & 0.045 &  & Blend 2.8\AA\ FWHM \\
4062.90 & \OII &  &  &  & \\
4068.60 & [\SII] + & 4068.61 & 16.29 & $\pm 0.13$ & \\
4069.62	& {\OII} +  &  &  &  & \\
4071.23	& {\OII} + & 4071.64 & 0.237 &  & \\
4072.16	& {\OII} +  &  &  &  & \\
4075.86	& \OII & 4074.15 & 0.215 &  & \\
4076.35 & [\SII] + & 4076.41 & 5.487 & $\pm 0.15$ & \\
4078.84	& \OII &  &  &  & \\
4083.90 & {\OII} + & 4084.46 & 0.041 &  & \\
4085.06	& \OII &  &  &  & \\
4089.29	& \OII & 4089.00 & 0.060 & $\pm 0.005$ & \\
4092.93 & \OII & 4093.80 & 0.033 &  & \\
4097.33 & {\NIII} + & 4097.30 & 3.852 &  & \\
4097.25 & {\OII} + &  &  &  & \\
4097.26 & {\OII} + &  &  &  & \\
4098.24 & {\OII} + &  &  &  & \\
4100.04 & {\HeII} (4-12) + &  &  &  & \\
4101.73 & H$\delta$ & 4101.76 & 29.27 & $\pm 0.8$ & \\
4120.82 & {\HeI} + & 4121.16 & 0.479 & $\pm 0.08$ & \\
4121.3	& \NII &  &  &  & \\
4123.46	& \NII ? & 4123.02 & 0.355 &  & \\
4129.32 & \OII & 4129.04 & 0.034 &  & \\
4132.80	& \OII & 4132.86 & 0.061 &  & \\
4144.32	&[\FeIII] + & 4144.20 & 0.501 &  & \\
4143.76	& \HeI &  &  &  & \\
4153.30 & \OII & 4153.15 & 0.051 & $\pm 0.013$ & Blend 3.2\AA\ FWHM \\
4157.75 & [\FII] + & 4156.56 & 0.084 & $\pm 0.008$ & \\
4156.53	& \OII    &  &  &  & \\
4163.33	& [\KV] & 4163.46 & 0.584 & $\pm 0.008$ & \\
4168.97	& {\HeI} + & 4169.28 & 0.110 & $\pm 0.008$ & \\
4169.22 & \OII &  &  &  & \\
4176.16 & \NII & 4175.76 & 0.055 & $\pm 0.008$ & \\
& ? & 4178.32 & 0.035 &  & \\
4180.9 & [\FeV] & 4181.00 & 0.066 & $\pm 0.008$ & \\
4186.8 & \CIII & 4186.81 & 0.141 & $\pm 0.013$ & Blend 2.9\AA\ FWHM \\
4189.7 & \OII & 4189.94 & 0.071 &  & \\
4195.6 & \NIII & 4195.63 & 0.158 &  & \\
4199.83 & {\HeII} (4-11) & 4199.79 & 1.859 &  & \\
4227.5 & [\NiIII] & 4228.00 & 0.202 &  & \\
4241.48 & [\MnIII] + & 4241.31 & 0.047 &  & \\
4241.79 & \NII &  &  &  & \\
& ? & 4255.91 & 0.156 &  & \\
4267.13 & \CII & 4267.00 & 0.080 & $\pm 0.008$ & \\
4273.06 & \OII & 4272.96 & 0.077 &  & \\
4275.5 & {\OII} + & 4276.24 & 0.057 & & Blend 2.5\AA\ FWHM \\
4276.7 & \OII &  &  &  & \\
4282.91 & {\OII} +  & 4283.05 & 0.043 &  & \\
4283.68 & \OII &  &  &  & \\
4287.39	& [\FeII] & 4287.41 & 0.041 &  & \\
4294.76 & \OII & 4294.12 & 0.063 & & Blend 3.6\AA\ FWHM \\
4317.14 & {\OII} + & 4318.55 & 0.048 &  & \\
4319.63 & \OII &  &  &  & \\
4331 & He(2-10)  & 4331.30 & 0.070 & & Broad He Raman Feature: \emph{see} \\
& (Raman) & & & & Pequinot \emph{et al.,} A\&A 1997, 323, 217. \\
4338.67	& {\HeII} (4-10) + &  &  &  & \\
4340.46 & H$\gamma$ & 4340.45 & 43.74 & $\pm 1.2$ & \\
4345.56 & \OII & 4345.26 & (0.15) & & Uncertain: on strong line wing \\
4359.34 & [\FeII] & 4359.28 & (0.14) & & Uncertain: on strong line wing \\
4363.21 & [\OIII] & 4363.23 & 38.18 & & \\
4366.89 & \OII & 4367.96 & (0.16) & & Uncertain: on strong line wing \\
& ? & 4376.48 & 0.066 &  & \\
4379.2 & \NIII & 4378.87 & 0.270 & & \\
4387.8 & \HeI & 4387.83 & 0.643 & & \\
4400 - & {\OI} + & 4413.77 & 0.554 & & Blend of many lines, \\
4417 & \NeII &  &  &  & 11.8\AA\ FWHM \\
4431.82 & \NII & 4431.62 & 0.164 &  & \\
4437.55 & \HeI & 4437.53 & 0.111 & $\pm 0.008$ & \\
4452.37 & {\OII} + ? & 4453.07 & 0.152 & & Blend 4.3\AA\ FWHM \\
4471.47 & {\HeI} & 	4471.45 & 5.428 & $\pm 0.1$ & \\
4491.2 & \OII & 4491.11 & 0.055 &  & \\
4492.64 & [\FeII] & 4493.09 & 0.058 &  & \\
4498.04	& [\MnIV] & 4498.60 & 0.083 & $\pm 0.02$ & \\
4510.92 & [\KIV] + & 4510.72 & 0.208 & $\pm 0.02$ & \\
4510.91 & \NIII  &  &  &  & \\
4514.6 & \NIII & 4514.74 & 0.097 & $\pm 0.025$ & \\
4518.15 & \NIII & 4518.32 & 0.082 & $\pm 0.015$ & \\
4519.62	& \NII & 4519.69 & 0.119 &  & \\
4519.63	& \OIII & 4519.48 & 0.105 &  & \\
4518 -  & \NIII , \CIII & 4522.86 & 0.128 & & Blend of {\NIII},{\CIII}, \\
4525 &  &  &  &  & 5\AA\ FWHM\\
4523.6 & \NIII & 4523.57 & 0.071 & $\pm 0.025$ & \\
4529.09	& [\MnIV] & 4529.23 & 0.080 &  & \\
4530.42 & \NII & 4530.30 & 0.104 & $\pm 0.035$ & \\
4534.57 & \NIII & 4534.56 & 0.071 & $\pm 0.025$ & \\
4541.59 & \HeII (4-9) & 4541.53 & 2.540 & $\pm 0.035$ & \\
4549.04	& [\MnIV] & 4549.53 & 0.064 & $\pm 0.035$ & \\
4552.53 & \NII & 4552.54 & 0.033 &  & \\
4554.0	& \BaII & 4553.51 & 0.104 &  & \\
4563.85	& [\MnIV] & 4563.10 & 0.045 &  & \\
4566.60	& [\MnIII] & 4566.62 & 0.067 &  & \\
4571.1	& {\MgI}] & 4579.97 & 0.233 &  & \\
4591.66 & [\MnIV] & 4591.28 & 0.047 &  & \\
4596.18 & \OII & 4596.14 & 0.038 &  & \\
4603.74	& \NV & 4603.21 & 0.094 &  & \\
4607.06	& [\FeIII] + & 4606.59 & 0.686  & $\pm 0.02$ & \\
4607.16 & \NII  &  &  &  & \\
4609.66	& \OII & 4609.67 & 0.047 &  & \\
4611.59	& \OII & 4611.66 & 0.070 &  & \\
4613.87	& {\NII} + & 4613.75 & 0.101 &  &  \\
4613.68	& \OII & 4613.19 & 0.183 &  & \\
4615.65	& [\CoI] & 4615.64 & 0.102 &  & \\
4619.97	& \NV & 4619.57 & 0.062  &  & \\
4624.92	& [\ArV] & 4624.92 & 0.397 & $\pm 0.020$ & \\
4629.39	& [\FeIII] & 4629.54 & 0.056  &  & \\
4634.14	& \NIII & 4634.10 & 2.926  &  & \\
4640.64	& \NIII & 4640.49 & 6.112 & $\pm 0.022$ & \\
4644.1 & {\CIII} + & 4646.42 & 0.110 & & Blend \\
4646.93	& {\NII} + &  &  &  & \\
4647.42 & {\CIII} + &  &  &  & \\
4647.80 & {\OII} + &  &  &  & \\
4649.13	& \OII & 4649.39 & 0.232 &  & \\
4658.10	& [\FeIII] & 4657.87 & 0.418 & $\pm 0.015$ & \\
4661.63	& \OII & 4661.76 & 0.118 &  & \\
4676.26	& \OII & 4675.83 & 0.091 &  & \\
4685.71	& {\HeII} (3-4) & 4685.82 & 67.49 & $\pm 0.15$ & \\
4701.62	& [\FeIII] & 4701.61 & 0.098 & $\pm 0.02$ & \\
& ? & 4707.31 & 0.125 &  & \\
4711.37 & [\ArIV] + & 4711.35 & 12.04 & $\pm 0.10$ & \\
4711.9 & [\NeIV] &  &  &  & \\
4713.14 & \HeI & 4713.51 & 3.020 & $\pm 0.10$ & \\
4724.15 & [\NeIV] + & 4724.82 & 3.686 & $\pm 0.13$ & \\
4725.62 & [\NeIV] &  &  &  & \\
4733.90 & [\FeIII] & 4733.97 & 0.057 & $\pm 0.02$ & \\
4740.16 & [\ArIV] & 4740.20 & 18.75 & $\pm 0.10$ & \\
4754.80 & [\FeIII] & 4754.47 & 0.070 & $\pm 0.008$ & \\
4769.40	& [\FeIII] & 4769.40 & 0.050 & $\pm 0.008$ & \\
4777.7 & [\FeIII] & 4777.86 & 0.025 &  & \\
4788.13 & \NII & 4788.32 & 0.052 &  & \\
4803.29 & \NII & 4802.91 & 0.070 &  & \\
4814.55 & [\FeII] & 4814.48 & 0.050 &  & \\
4852 & He(2-8) & 4852.22 & 0.470 & & He Raman Feature: 5\AA\ FWHM \emph{see} \\
& (Raman) & & & &  Pequinot \emph{et al.} A\&A 1997, 323, 217. \\
4859.32	& {\HeII} (4-8) + &  &  &  & \\
4861.32 & H$\beta$  & 4861.30 & 100.0  & $\pm 0.3$ & Error:
measurement error only.\\
& & & & & (\emph{for systematic errors, see text.}) \\
4889.6 & [\FeII] & 4889.55 & 0.031 & $\pm 0.005$ & \\
& ? & 4893.60 & 0.059 &  & \\
& ? & 4902.62 & 0.052 &  & \\
& ? & 4906.49 & 0.071 &  & \\
4921.93 & \HeI & 4921.94 & 1.562 &  & \\
4931.23 & [\OIII] & 4930.83 & 0.381 &  & \\
& ? & 4938.64 & 0.075 &  & Blend? \\
& ? & 4944.60 & 0.314 &  & \\
4958.91 & [\OIII] & 4958.86 & 380.8 & $\pm 1$ & \\
4987.20 & [\FeIII] & 4988.76 & 0.123 & $\pm 0.01$ & \\
4994.36 & \NII & 4994.84 & 0.077 &  & \\
5006.73 & [\OIII] & 5006.73 & 1055 & $\pm 3$ & \\
5015.68 & \HeI & 5016.33 & (2.0) &  & Measurement difficult; on line wing. \\
5032.43	& \SII & 5032.60 & 0.056 & $\pm 0.005$ & \\
5041.03	& \SiII & 5041.00 &	1.704 &  & \\
5047.74 & \HeI & 5047.9	& 0.250 &  & Wavelength scale suspect 5040-5180. \\
5055.98 & {\SiII} + & 5056.20 & 0.758 &  & \\
5056.31 & {\SiII} &  &  &  & \\
& ? & 5074.5 & 0.012 &  & \\
5084.8 & [\FeIII] & 5086.9 & 0.037 & $\pm 0.005$ & \\
5103.30 & \SII & 5103.1 & 0.033 &  & \\
5111.63	& [\FeII] & 5112.2 & 0.028 & $\pm 0.006$ & \\
& ? & 5132.1 & 0.031 &  & \\
& ? & 5150.0 & 0.053 &  & \\
5158.77 & [\FeII] & 5157.6 & 0.116 &  & \\
5176.4 & [\FeVI] & 5176.0 & 0.242 &  & \\
5191.82 & [\ArIII] & 5191.80 & 0.266 & $\pm 0.02$ & \\
5197.90 & [\NI] & 5197.88 & 4.457 & $\pm 0.06$ & \\
5200.26 & [\NI] & 5200.88 & 3.310 &  & \\
5220.06 & [\FeII] & 5219.61 & 0.020 & $\pm 0.005$ & \\
& ? & 5233.60 & 0.015 &  & \\
5261.62 & [\FeII] & 5261.42 & 0.105 &  & \\
5270.40	& [\FeIII] & 5270.38 & 0.107 &  & \\
5273.38 & [\FeII] & 5273.24 & 0.024 & $\pm 0.004$ & \\
5277.8 & [\FeVI] & 5276.95 & 0.093 & $\pm 0.006$ & \\
5289.79 & [\FeVI] & 5290.44 & 0.219 & $\pm 0.008$ & \\
5296.82 & [\FeII] + & 5298.63 & 0.037 &  & \\
5298.87	& [\FeII] &  &  &  & \\
& ? & 5304.51 & 0.015 &  & \\
5309.11 & [\CaV] & 5309.26 & 0.310	& $\pm 0.01$ & \\
5323.30 & [\ClIV] & 5323.15 & 0.188 &  & \\
5335.18 & [\FeVI] & 5335.25 & 0.107 & & Blend 3.0\AA\ FWHM \\
5346 & [\KrIV] & 5346.10 & 0.020 &  & \\
& ? & 5359.40 & 0.013 &  & \\
5364 & [\RbV] & 5363.24 & 0.012 &  & \\
& ? & 5371.29 & 0.016 &  & \\
5376.45 & [\FeII] & 5376.37 & 0.015 & $\pm 0.002$ & \\
5411.53 & \HeII (4-7) & 5411.49 & 7.327 & $\pm 0.01$ & \\
5423.9 & [\FeVI] & 5424.37 & 0.049 &  & \\
5426.6 & [\FeVI] & 5426.62 & 0.017 &  & \\
5433.13 & [\FeII] & 5432.75 & 0.004 &  & \\
5460.69 & [\CaVI] & 5460.64 & 0.024 & $\pm 0.002$ & \\
& ? & 5467.31 & 0.027 & $\pm 0.002$ & \\
& ? & 5470.24 & 0.010 &  & \\
5484.9 & [\FeVI] & 5484.81 & 0.031 & $\pm 0.002$ & \\
& ? & 5494.30 & 0.004 &  & \\
5495.70 & {\NII} +  & 5495.39 & 0.005 &  & \\
5495.72	& [\FeII] &  &  &  & \\
& ? & 5506.92 & 0.006 &  & \\
5517.66 & [\ClIII] & 5517.54 & 0.484 & $\pm 0.004$ & \\
5527.33 & [\FeII] & 5526.95 & 0.031 & $\pm 0.002$ & \\
5530.24	& \NII & 5530.16 & 0.019 & $\pm 0.003$ & \\
5537.60 & [\ClIII] & 5537.71 & 1.099 & $\pm 0.004$ & \\
5543.81	& \CI & 5543.94 & 0.015 & $\pm 0.003$ & \\
5551.95 & \NII & 5551.88 & 0.025 & $\pm 0.003$ & \\
5555.03	& \OI & 5555.90	& 0.006 & $\pm 0.002$ & \\
5568.35	& \SiII & 5568.42 & 0.009 &  & \\
5577.34 & [\OI] & 5577.30 & 0.279 & $\pm 0.006$ & \\
5592.37	& \OIII & 5592.12 & 0.074 & $\pm 0.005$ & \\
& ? & 5597.57 & 0.009 & $\pm 0.001$ & \\
5602.3 & [\KVI] & 5602.04 & 0.110 & $\pm 0.004$ & \\
& ? & 5618.70 & 0.032 &  & \\
& ? & 5622.10 & 0.040 &  & \\
5631.07 & [\FeVI] & 5630.99 & 0.052 &  & \\
5644.80	& [\FeIV] & 5645.02 & 0.007 &  & \\
5666.63 & \NII & 5666.65 & 0.089 & $\pm 0.008$ & \\
5676.02 & {\NII} + & 5676.67 & 0.061 &  & \\
5677 & [\FeVI] &  &  &  & \\
5679.56 & \NII & 5679.50 & 0.154 &  & \\
5686.21 & \NII & 5686.07 & 0.017 &  & \\
5692.04 & [\FeIV] + & 5692.07 & 0.035 &  & \\
5693.56	& [\MnV] &  &  &  & \\
5703.4 & [\MnV] & 5702.08 & 0.063 & $\pm 0.001$ & \\
5710.77 & \NII & 5710.71 & 0.023 & $\pm 0.001$ & \\
5721.1 & [\FeVII] & 5721.19 & 0.277 & $\pm 0.001$ & \\
5739.73 & \SiIII & 5739.42 & 0.012 & $\pm 0.001$ & \\
5754.60	& [\NII] & 5754.81 & 23.18 & $\pm 0.05$ & \\
5784.94	& {\HeII} (5-40) & 5785.15 & 0.027 &  & \\
5801.51 & \CIV +  & 5800.65 & 0.124 &  & \\
5800.48	& {\HeII} (5-37) &  &  &  & \\
5806.57	& {\HeII} (5-36) & 5806.27 & 0.050 &  & \\
5812.14	& {\CIV} + & 5812.52 & 0.058 & & Blend 2.50\AA\ FWHM \\
5813.19	& {\HeII} (5-35) &  &  &  & \\
5820.43	& {\HeII} (5-34) & 5820.31 & 0.046 &  & \\
5828.36	& {\HeII} (5-33) & 5828.21 & 0.038 &  & \\
5837.06	& {\HeII} (5-32) & 5836.88 & 0.050 &  & \\
5847.25 & {\HeII} (5-31) & 5846.58 & 0.046 &  & \\
& ? & 5852.73 & 0.061 &  & \\
5857.26	& {\HeII} (5-30) & 5857.23 & 0.064 & $\pm 0.01$ & \\
& ? & 5861.39 & 0.078 &  & \\
5862.6 & [\MnV] + & 5864.10 & 0.009 & $\pm 0.003$ & \\
5869.02	& {\HeII} (5-29) & 5869.09 & 0.080 & $\pm 0.01$ & \\
5875.66 & He I &  5875.65 & 20.35 & $\pm 0.4$ & \\
5882.12	& {\HeII} (5-28) & 5881.99 & 0.051 &  & \\
5896.78	& {\HeII} (5-27) & 5895.55 & 0.064 & $\pm 0.01$ & \\
5913.24	& {\HeII} (5-26) & 5913.27 & 0.085 & $\pm 0.006$ & \\
& ? & 5921.97 & 0.006 &  & \\
5927.81 & \NII & 5927.66 & 0.011 & $\pm 0.002$ & \\
5931.78	& {\NII} + & 5931.89 & 0.091 & $\pm 0.002$ & \\
5931.83	& {\HeII} (5-25) &  &  &  & \\
5941.65 & \NII & 5941.23 & 0.022 & $\pm 0.001$ & \\
& ? & 5945.27 & 0.012 & $\pm 0.002$ & Possible blend? \\
5952.93	& {\HeII} (5-24) & 5952.90 & 0.104 &  & \\
5957.56	& \SiII & 5957.60 & 0.011 & & Blend 4\AA\ FWHM \\
& ? & 5961.69 & 0.031 &  & \\
& ? & 5969.23 & 0.009 &  & \\
5977.03	& {\HeII} (5-23) & 5977.18 & 0.122 &  & \\
5978.93	& \SiII & 5977.08 & 0.119 & $\pm 0.002$ & \\
& ? & 5980.58 & 0.012 &  & \\
& ? & 5989.55 & 0.014 & $\pm 0.002$ & \\
6004.72 & {\HeII} (5-22) & 6004.66 & 0.118 &  & \\
6024.40 & [\MnV] & 6024.81 & 0.015 &  & \\
6036.78 & {\HeII} (5-21) & 6037.19 & 0.139 &  & \\
6074.19 & {\HeII} (5-20) & 6074.20 & 0.174 &  & \\
6084.9 & [\MnV] & 6083.64 & 0.044 &  & \\
6086.40 & [\CaV] & 6086.64 & 0.464 &  & \\
6101.8 & [\KIV] & 6101.33 & 0.756 & $\pm 0.002$ & \\
6118.26 & {\HeII} (5-19) & 6118.26 & 0.179 & $\pm 0.002$ & \\
6131 & [\BrIII] & 6130.53 & 0.004 &  & \\
& ? & 6134.47 & 0.007 & $\pm 0.001$ & Blend 2.7\AA\ FWHM \\
6141.7 & \BaII & 6141.66 & 0.008 &  & \\
6151.43 & \CII & 6150.95 & 0.009 &  & \\
6157.6 & [\MnV] & 6157.44 & 0.036 & $\pm 0.002$ & \\
6161.83	& [\ClII] & 6161.8 & 0.011 &  & \\
& ? & 6165.75 & 0.029 &  & \\
6167.7 & [\MnV] & 6167.70 & 0.008 &  & \\
6170.69 & {\HeII} (5-18) & 6170.67 & 0.212 &  & \\
& ? & 6198.31 & 0.022  &  & \\
& ? & 6200.06 & 0.048 & & Blend 3.3\AA\ FWHM \\
6218.4 & [\MnV] & 6218.88 & 0.026 & $\pm 0.002$ & \\
6219.2 & [\MnV] & 6221.58 & 0.026 & $\pm 0.002$ & \\
6228.6 & [\KVI] & 6228.26 & 0.200 &  & \\
6233.82	& {\HeII} (5-17) & 6233.78 & 0.251 &  & \\
& ? & 6273.19 & 0.011 &  & \\
& ? & 6277.89 & 0.114 &  & \\
& ? & 6289.59 & 0.018 & $\pm 0.002$ & \\
6300.30 & [\OI] & 6300.40 & 22.31 & $\pm 0.4$ & \\
6312.10 & [\SIII] & 6312.06 & 6.581 & $\pm 0.15$ & \\
& ? & 6341.30 & 0.028 &  & \\
6345.4 & [\MnV]? & 6343.55 & 0.055 & $\pm 0.006$ & \\
6347.09 & \SiII & 6347.18 & 0.398 &  & \\
6363.78 & [\OI] & 6363.70 & 7.839 & $\pm 0.15$ & \\
6371.36 & \SiII & 6371.27 & 0.893 &  & \\
& ? & 6383.70 & 0.005 &  & \\
6393.7 & [\MnV] & 6393.55 & 0.068 &  & \\
& ? & 6402.28 & 0.001 &  & \\
6406.38 & {\HeII} (5-15) & 6406.17 & 0.365 &  & \\
& ? & 6412.21 & 0.052 &  & \\
6427.1 & [\CaV] & 6426.87 & 0.004 & $\pm 0.001$ & \\
6434.73 & [\ArV] & 6434.76 & 5.314 &  & \\
& ? & 6444.23 & 0.006 & & Blend 3.6\AA\ FWHM \\
& ? & 6451.69 & 0.011 & $\pm 0.001$ & \\
& ? & 6455.09 & 0.007 & $\pm 0.001$ & Blend 2.5\AA\ FWHM \\
& ? & 6460.81 & 0.031 & $\pm 0.003$ & Blend 2.8\AA\ FWHM \\
6465.95 & {\SiI}? & 6466.02 & 0.032 & $\pm 0.003$ & \\
6473.86 & [\FeII] & 6473.83 & 0.023 &  & \\
& ? & 6478.04 & 0.069 & & Blend 2.7\AA\ FWHM \\
6482.05 & \NII & 6482.05 & 0.025 &  & \\
6496.9 & \BaII & 6496.37 & 0.015 & & Blend 3.3\AA\ FWHM \\
6500.04 & [\CrIII]? & 6500.27 & 0.076 & $\pm 0.004$ & \\
6516.53 & [\VI] & 6516.40 & 0.043 & $\pm 0.002$ & \\
& ? & 6521.68 & 0.006 &  & \\
6527.10 &{\HeII} (5-14) + & 6526.48 & 1.003	& $\pm 0.001$ & Blend
3.3\AA\ FWHM \\
6527.24 & [\NII] &  &  &  & \\
6548.04 & [\NII] & 6548.09 & 173.1 & $\pm 0.8$ & \\
6560.2 & \HeII & 6559.98 & 9.463 &  & Uncertain, on bright line wing\\
&  &  &  &  & \\
6562.80 & H$\alpha$ & 6562.78 & 292.5 & $\pm 0.6 $ & \\
&  &  &  &  & \\
& ? & 6575.05 & 0.088 &  & \\
6583.46 & [\NII] & 6583.46	& 504.6 & $\pm 0.8 $ & \\
& ? & 6611.00 & 0.011 &  & \\
& ? & 6624.72 & 0.013 &  & \\
6655.52 & \CI & 6655.71 & 0.045 &  & \\
6666.66 & {\OII} + & 6666.98 & 0.018 &  & Blend 3.8\AA\ FWHM \\
6666.80	& [\NiII] + &  &  &  & \\
6667.01	& [\FeII] &  &  &  & \\
6678.15 & \HeI & 6678.20 & 4.328& $\pm 0.004$ & \\
6683.20 & {\HeII} (5-13) & 6683.20 & 0.586 & $\pm 0.002$ & \\
6693.96	& \CI] & 6693.95 & 0.112 &  & \\
& ? & 6707.56 & 0.132 &  & \\
6709.64 & [\LiI]? & 6710.07 & 0.155 & $\pm 0.006$ & \\
6716.44 & [\SII] & 6716.43 & 12.75 & $\pm 0.01 $ & \\
6730.81 & [\SII] & 6730.79 & 23.98 & $\pm 0.01 $ & \\
6744.1 & {\HeI} +  & 6746.16 & 0.056 &  & Blend 4.3\AA\ FWHM \\
6746.3 & {\CIV} &  &  &  & \\
6746.7 & {\CIV} &  &  &  & \\
6747.5 & {\CIV} &  &  &  & \\
6795.1 & [\KIV] & 6795.22 & 0.188 & $\pm 0.002$ & \\
6830 & {\OVI} & 6829.64 & 0.300 &  & Raman line with velocity structure: \\
& (Raman)  &  &  &  & 8.3 \AA\ FWHM \\
6850.33 & [\MnII] & 6850.19 & 0.084 &  & \\
6855.88 & \HeI & 6855.85 & 0.018 & $\pm 0.002$ & \\
& ? & 6867.56 & 0.027 &  & \\
6890.90 & {\HeII} (5-12) & 6891.00 & 0.661 &  & \\
6927.85 & \SII & 6928.23 & 	0.046 & $\pm 0.002$ &  \\
6984.08	& [\FeII] & 6984.36 & 0.022 &  & \\
6989.45 & \HeI & 6989.43 & 0.011 &  & \\
7005.4 & [\ArV] & 7005.63 & 10.70 & $\pm 0.01$ & \\
7046.88 & \SiI & 7046.81 & 0.029 &  & \\
& ? & 7057.9 & 0.074 &  & Blend 4.5\AA\ FWHM \\
7065.19 & \HeI & 7065.23 & 10.29 & $\pm 0.01$ &  \\
7082.1 & \SiI & 7082.06 & 0.009 &  & \\
7087 & {\OVI} & 7087.3 & 0.022 &  & Raman line with velocity structure: \\
& (Raman)  &  &  &  & 9 \AA\ FWHM \\
& ? & 7114.41 & 0.096 &  & \\
7135.8 & [\ArIII] & 7135.76 & 26.44 & $\pm 0.05$ & \\
7155.16 & [\FeII] & 7154.98 & 0.208 & $\pm 0.003$ & \\
7160.58 & \HeI & 7160.62 & 0.028 & $\pm 0.005$ & \\
7170.5 & [\ArIV] & 7170.61 & 1.559 & $\pm 0.015$ & \\
7177.52 & {\HeII} (5-11) & 7177.60 & 0.882 & $\pm 0.004$ & \\
7237.40	& [\ArIV] & 7237.70 & 1.187 & $\pm 0.006$ & \\
7252.30	& \SiI & 7252.49 & 0.016 & $\pm 0.001$ & \\
7255.8 & [\NiII] & 7255.98 & 0.016 &  & \\
7262.7 & [\ArIV] & 7262.87 & 1.268 & $\pm 0.008$ & \\
7281.35 & \HeI & 7281.34 & 1.217 & $\pm 0.01$ & \\
7291.47	& [\CaII] & 7290.83 & 0.037 &  & \\
7298.03 &  \HeI	 & 7298.00 & 0.054 & $\pm 0.001$ & \\
7306.85 & {\OIII} + & 7307.18 & 0.051 & $\pm 0.004$ & \\
7307.12 & \OIII &  &  &  & \\
7318.92 & [\OII] + & 7320.06 & 10.51 &  & \\
7319.99	& [\OII] &  &  &  & \\
7323.89	& [\CaII] & 7323.60 & 0.049 &  & \\
7329.66	& [\OII] +  & 7330.37 & 9.182 &  & \\
7330.73	& [\OII] &  &  &  & \\
7377.83 & [\NiII] & 7377.54	& 0.111 & $\pm 0.005$ & \\
7388.2 & [\FeII] & 7387.76 & 0.049 & & Possibly a blend. \\
7411.61 & [\NiII] & 7411.32	& 0.010 &  &\\
7423.61	& \NI & 7423.40 & 0.008  &  & \\
& ? & 7439.90 & 0.015 & $\pm 0.003$ & \\
7442.30 & \NI & 7442.53 & 0.006  &  & \\
7448.26	& \NI & 7447.91 & 0.005  &  & \\
7452.54 & [\FeII] & 7452.29 & 0.073 & $\pm 0.001$ & \\
7468.31 & \NI & 7468.50 & 0.009 & $\pm 0.001$ & \\
7487.04 & [\FeII] + & 7487.19 & 0.007 &  & Blend 2.6\AA\ FWHM \\
7486.7	& \CIII &  &  &  & \\
7499.85	& \HeI & 7499.82 & 0.054 & $\pm 0.001$ & \\
7530.0 & [\ClIV] & 7530.38 & 0.848 & $\pm 0.001$ & \\
7561.42 & [\MnII] & 7561.55 & 0.010 & $\pm 0.001$ & \\
7578.80	& \SiI & 7578.80 & 0.046  &  & \\
7581.5 & \NIV & 7581.72 & 0.133 & $\pm 0.02$ & Blend 4.2\AA\ FWHM \\
7592.75 & {\HeII} (5-10) + & 7592.76 & 1.487 & $\pm 0.01$ & \\
7592.0 & \CIV  & 7592.76 & 1.497 & $\pm 0.03$ & \\
7686.82 & \NIII & 7686.67 & 0.025  &  & \\
7686.94	& [\FeII] &  &  &  & \\
7703.0 & {\NIV} +  & 7703.24 & 0.242  &  & \\
7703.4 & \NII &  &  &  & \\
& ? & 7712.67 & 0.038  &  & \\
& ? & 7716.78 & 0.090  &  & \\
7726.2 & \CIV & 7726.11 & 0.028  &  & \\
& ? & 7731.02 & 0.027  &  & \\
& ? & 7737.88 & 0.035  &  & \\
7751.1 & [\ArIII] & 7751.10 & 7.384  &  & \\
7816.13 & \HeI & 7816.15 & 0.077  &  & \\
7837.76	& \ArII  & 7837.72 & 0.011  &  & \\
& ? & 7857.56 & 0.0126 & $\pm 0.003$ & \\
7860.8 & \CIV  & 7861.37 & 0.019 & $\pm 0.002$ & \\
7875.99 & [\PII] + & 7875.98 & 0.040 & $\pm 0.001$ & \\
7876.6 & \CIV  &  &  &  & \\
& ? & 7883.45 & 0.039  &  & \\
7924.2 & [\FeIII] +	& 7924.66 & 0.0311  &  & \\
7924.8 & [\FeIII] &  &  &  & \\
& ? & 7935.38 & 0.036 &  & \\
& ? & 7968.65 & 0.017 &  & \\
7999.4 	& {\HeI} + & 8000.10 & 0.0445  &  & \\
8000.08 & [\CrII] &  &  &  & \\
8015.0 & \CI &  8016.10 & 0.027 &  & \\
8015.8 & {\HeI} (4-20) &  &  &  & \\
8018.57 & \CI & 8018.88 & 0.024 &  & \\
8021.25	& \CI & 8021.40 & 0.025 &  & \\
& ? & 8025.56 & 0.005 &  & \\
8034.8 & {\HeI} (4-19) & 8035.52 & 0.006 &  & \\
8039.77	& \CI & 8039.50 & 0.056 &  & \\
8046.3 & [\ClIV] & 8045.62 & 1.931 & $\pm 0.002$ & \\
8057.3 & {\HeI}  (4-18) &  8057.61 & 0.014 &  & \\
8064.8 & \NII & 8064.78 & 0.081 &  & \\
8083.8 & \CI & 8084.00 & 0.013 & $\pm 0.002$ & \\
8116.49 & \OII & 8116.32 & 0.015 &  & \\
8125.30	& [\CrII] & 8125.39 & 0.032 &  & \\
& ? & 8137.36 & 0.0154 &  & \\
& ? & 8160.12 & 0.060 &  & \\
& ? & 8196.55 & 0.121 & $\pm 0.008$ & \\
& ? &  8216.50 & 0.016 &  & \\
8229.67 & [\CrII] & 8230.00 & 0.032 & $\pm 0.013$ & \\
8236.79 & {\HeII} (5-9) & 8236.75 & 2.152 & $\pm 0.1$ & \\
8267.94	& {\HI} (P34) & 8267.93 &  &  & \\
8271.93	& {\HI} (P33) & 8271.92 &  &  & \\
8276.31 & {\HI} (P32) & 8276.41 & 0.094 & $\pm 0.008$ & \\
8281.12 & {\HI} (P31) & 8281.33 & 0.134 & $\pm 0.008$ & \\
8286.43 & {\HI} (P30) & 8286.42	& 0.122 & $\pm 0.013$ & \\
8292.31 & {\HI} (P29) & 8292.18 & 0.119 & $\pm 0.008$ & \\
8298.83 & {\HI} (P28) &	8298.84 & 0.139 & $\pm 0.007$ & \\
& ? & 8303.13 & 0.014 &  & \\
8306.11 & {\HI} (P27) & 8306.42 & 0.160 & $\pm 0.008$ & Blend 2.8\AA\ FWHM \\
8314.26 & {\HI} (P26) & 8314.20 & 0.151 & $\pm 0.009$ & \\
& ? & 8320.17 & 0.032 &  & \\
8323.42 & {\HI} (P25) & 8323.40 & 0.158 & $\pm 0.006$ & \\
& ? & 8329.77 & 0.031 & $\pm 0.006$ & \\
8333.78 & {\HI} (P24) & 8333.84 & 0.190 & $\pm 0.01$ & \\
8342.35	& \HeI & 8342.38 & 0.048 & $\pm 0.008$ & \\
8345.55 & {\HI} (P23) & 8345.53 & 0.176 & $\pm 0.013$ & \\
& ? & 8348.55 & 0.013  &  & \\
& ? & 8355.45 & 0.011 &  & \\
8359.00 & {\HI} (P22) & 8359.06 & 0.226 & $\pm 0.013$ & \\
8361.73 & \HeI & 8361.73 & 0.161 & $\pm 0.02$ & \\
& ? & 8370.77 & 0.020  & $\pm 0.004$ & \\
8374.47 & {\HI} (P21) & 8374.51 & 0.220 & $\pm 0.006$ & \\
& ? & 8379.29 & 0.025  &  & \\
& ? & 8386.57 & 0.072 &  & \\
& ? & 8388.13 & 0.030  &  & \\
8392.40 & {\HI} (P20) & 8392.40 & 0.254  & $\pm 0.015$ & \\
8399 & \HeII & 8398.61 & 0.036 & $\pm 0.008$ & Blend 3.7\AA\ FWHM \\
& ? & 8409.82 & 0.014  &  & \\
8413.32 & {\HI} (P19) & 8413.33 & 0.271 & $\pm 0.003$ & \\
& ? & 8421.81 & 0.030 & $\pm 0.003$ & \\
& ? & 8424.61 & 0.010 &  & \\
8434.0 & [\ClIII] & 8433.87 & 0.092 &  & \\
8437.95 & {\HI} (P18) & 8437.95 & 0.313 &  & \\
8446.4 & \OI & 8446.58 & 0.190 &  & Possible Blend \\
& ? & 8451.16 & 0.020 &  & \\
& ? & 8453.73 & 0.013 &  & \\
& ? & 8459.16 & 0.017 &  & \\
& ? & 8463.80 & 0.029 &  & \\
8467.25 & {\HI} (P17) & 8467.26 & 0.364 & $\pm 0.003$ & \\
& ? & 8474.35 & 0.009 &  & Blend 3.3\AA\ FWHM \\
8480.68	& {\HeI} + & 8480.84 & 0.092 & $\pm 0.003$ & \\
8481.2 & [\ClIII] &  &  &  & \\
8486.2 & \HeI & 8486.30 & 0.023 & $\pm 0.003$ & \\
8488.7 & \HeI & 8488.73 & 0.012  &  & \\
8500.2 & [\ClIII] & 8499.67 & 0.101  &  & \\
8502.48 & {\HI} (P16) & 8502.35 & 0.458 & $\pm 0.03$ & \\
8519.3	& {\HeII} (6-31) & 8519.14 & 0.033& $\pm 0.003$ & \\
8528.9	& \HeI & 8529.09 & 0.024 & $\pm 0.003$ & \\
8532.1 	& \HeI & 8531.91 & 0.008  &  & \\

\end{supertabular}
\normalsize
\newpage

\begin{table}
\centering
\begin{tabular}{|c|c|c|c|c|}
\hline
~Ion~ & Line (\AA) & Recombination &Intensity& Abundance\\
~&~&Coefficient$^1$&~&~\\
\hline
H$^+$    &4861  &2.10E-14&100.0&1.00\\
He$^+$   &5876  &3.11E-14&20.4 &0.16\\
~        &4471  &8.93E-15&5.43 &0.11\\          
~        &4922  &2.62E-15&1.56 &0.13 \\         
He$^{+2}$&4686  &2.36E-13&67.5 &0.058\\
C$^{+2}$ &4267  &1.77E-13&0.080&8.3E-5\\
C$^{+3}$ &8197  &2.34E-13&0.121&1.9E-4\\
~	 &4187  &9.66E-14&0.141&2.6E-4\\
C$^{+4}$ &7726  &6.17E-13&0.028&1.5E-5\\
N$^{+2}$ &5941.6&2.70E-14&0.022&2.1E-4\\  
~	 &4040.9&5.70E-14&0.070&2.2E-4\\  
~	 &4239.4&3.70E-14&0.047&2.2E-4\\  
~	 &5679.6&5.80E-14&0.154&6.5E-4\\  
N$^{+3}$ &4379  &3.95E-13&0.270&1.3E-4\\
N$^{+4}$ &7703  &4.03E-13&0.242&2.0E-4\\  
~	 &7582  &1.57E-13&0.133&2.8E-4\\
O$^{+2}$ &4089.3&1.84E-14&0.060&5.7E-4\\
~	 &4132.8&1.07E-14&0.061&1.02E-3\\  
~	 &4649.1&1.04E-13&0.232&4.5E-4\\  
~	 &4676.2&2.03E-14&0.091&9.1E-4\\  
O$^{+3}$ &5592  &6.95E-15&0.074&2.5E-3\\  
O$^{+4}$ &7715  &4.03E-13&0.090&7.5E-5\\   
\hline
\end{tabular}
\caption{NGC 6302 Recombination line ion abundances. Some of the weaker lines
may be blends hence the large variation.  (1)Measured at 15,000K.}
\label{abund}
\end{table}
\end{document}